\documentclass[twocolumn,astrosymb]{aastex631}

\usepackage{amsmath}
\usepackage{url}
\usepackage{gensymb}

\usepackage{verbatim}

\newcommand{\Sg}{Sgr~A*}

\newcommand{\e}{\mathrm{e}}
\newcommand{\p}{\mathrm{p}}
\newcommand{\Msol}{\hbox{M$_\odot$}}

\newcommand{\0}{\phantom{0}}

\begin{document}


\title{	First mid-infrared detection and modeling of a flare from \Sg\
}
\author[0000-0002-9156-2249]{Sebastiano D. von Fellenberg}
\affiliation{Max Planck Institute for Radio Astronomy, Bonn \& 53121, Germany}
\correspondingauthor{Sebastiano D. von Fellenberg}
\email{sfellenberg@mpifr-bonn.mpg.de}

\author[0009-0003-9906-2745]{Tamojeet Roychowdhury}
\affiliation{Max Planck Institute for Radio Astronomy, Bonn \& 53121, Germany}
\affiliation{Department of Electrical Engineering, Indian Institute of Technology Bombay, Mumbai 400076, India }
\author[0000-0003-3503-3446]{Joseph M. Michail}
\altaffiliation{NSF Astronomy and Astrophysics Postdoctoral Fellow}
\affiliation{Center for Astrophysics $|$ Harvard \& Smithsonian, 60 Garden Street, Cambridge, MA 02138-1516; USA}
\author[0009-0004-8539-3516]{Zach Sumners}
\affiliation{McGill University, Montreal QC H3A 0G4, Canada}%
\affiliation{Trottier Space Institute, 3550 Rue University, Montréal, Québec, H3A 2A7, Canada}
\author[0000-0002-4463-2902]{Grace Sanger-Johnson}
\affiliation{Michigan State University, Department of Physics and Astronomy,  East Lansing, MI 48824, USA}
\author[0000-0002-0670-0708]{Giovanni G. Fazio}
\affiliation{Center for Astrophysics $|$ Harvard \& Smithsonian, 60 Garden Street, Cambridge, MA 02138-1516; USA}
\author[0000-0001-6803-2138]{Daryl Haggard}
\affiliation{McGill University, Montreal QC H3A 0G4, Canada}
\author[0000-0002-5599-4650]{Joseph L. Hora}
\affiliation{Center for Astrophysics $|$ Harvard \& Smithsonian, 60 Garden Street, Cambridge, MA 02138-1516; USA}
\author[0000-0001-7801-0362]{Alexander Philippov}
\affiliation{University of Maryland, College Park, MD 20742, USA.}
\author[0000-0002-7301-3908]{Bart Ripperda}
\affiliation{Canadian Institute for Theoretical Astrophysics, University of Toronto, 60 St. George Street, Toronto, ON M5S 3H8, Canada.}
\affiliation{Dunlap Institute for Astronomy and Astrophysics, University of Toronto, 50 St. George Street, Toronto, ON M5S 3H4, Canada.}
\affiliation{Department of Physics, University of Toronto, 60 St. George Street, Toronto, ON M5S 1A7, Canada.}
\affiliation{Perimeter Institute for Theoretical Physics, \\\small 31 Caroline Street North, Waterloo, ON N2L 2Y5, Canada.}
\author{Howard A. Smith}
\affiliation{Center for Astrophysics $|$ Harvard \& Smithsonian, 60 Garden Street, Cambridge, MA 02138-1516; USA}
\author[0000-0002-9895-5758]{S. P. Willner}
\affiliation{Center for Astrophysics $|$ Harvard \& Smithsonian, 60 Garden Street, Cambridge, MA 02138-1516; USA}
\author[0000-0003-2618-797X]{Gunther Witzel}
\affiliation{Max Planck Institute for Radio Astronomy, Bonn \& 53121, Germany}
\author[0000-0002-2967-790X]{Shuo Zhang}
\affiliation{Michigan State University, Department of Physics and Astronomy,  East Lansing, MI 48824, USA}
\author{Eric E. Becklin}
\affiliation{Department of Physics \& Astronomy,  University of California, Los Angeles, 90095-1547, USA}
\author[0000-0003-4056-9982]{Geoffrey C. Bower}
\affiliation{Academia Sinica Institute of Astronomy and Astrophysics, 645 N. A'ohoku Pl., Hilo, HI 96720, USA}
\author[0000-0002-8776-1835]{Sunil Chandra}
\affiliation{Physical Research Laboratory, Navrangpura, Ahmedabad, 380009}
\author[0000-0001-9554-6062]{Tuan Do}
\affiliation{Department of Physics \& Astronomy,  University of California, Los Angeles, 90095-1547, USA}
\author[0000-0003-4801-0489]{Macarena Garcia Marin}
\affiliation{European Space Agency (ESA), ESA Office, Space Telescope Science Institute, 3700 San Martin Drive, Baltimore, MD 21218, USA}
\author[0000-0003-0685-3621]{Mark A. Gurwell}
\affiliation{Center for Astrophysics $|$ Harvard \& Smithsonian, 60 Garden Street, Cambridge, MA 02138-1516; USA}
\author[0000-0001-8921-3624]{Nicole M. Ford}
\affiliation{McGill University, Montreal QC H3A 0G4, Canada}%
\affiliation{Trottier Space Institute, 3550 Rue University, Montréal, Québec, H3A 2A7, Canada}
\author[0000-0001-6906-772X]{Kazuhiro Hada}
\affiliation{Graduate School of Science, Nagoya City University,  Yamanohata 1, Mizuho-cho, Mizuho-ku, Nagoya, 467-8501, Aichi, Japan}
\author[0000-0001-9564-0876]{Sera Markoff}
\affiliation{Anton Pannekoek Institute for Astronomy, University of Amsterdam, Science Park 904, 1098 XH Amsterdam, The Netherlands}
\affiliation{Gravitation and Astroparticle Physics Amsterdam Institute, University of Amsterdam, Science Park 904, 1098 XH 195 196 Amsterdam, The Netherlands}
\author[0000-0002-6753-2066]{Mark R. Morris}
\affiliation{Department of Physics \& Astronomy,  University of California, Los Angeles, 90095-1547, USA}
\author[0000-0002-8247-786X]{Joey Neilsen}
\affiliation{Villanova University Department of Physics, 800 E. Lancaster Ave., Villanova PA, 19085, USA}
\author[0000-0001-7134-9005]{Nadeen B. Sabha}
\affiliation{Innsbruck, Institut für Astro- und Teilchenphysik, Technikerstr. 25/8, 6020 Innsbruck, Austria}
\author[0009-0001-1040-4784]{Braden Seefeldt-Gail}
\affiliation{Canadian Institute for Theoretical Astrophysics, University of Toronto, 60 St. George Street, Toronto, ON M5S 3H8, Canada.}
\affiliation{Dunlap Institute for Astronomy and Astrophysics, University of Toronto, 50 St. George Street, Toronto, ON M5S 3H4, Canada.}
\affiliation{Dunlap Institute for Astronomy \& Astrophysics, University of Toronto, 50 St. George Street, Toronto, ON M5S 3H4, Canada}
\begin{abstract}
The time-variable emission from the accretion flow of \Sg, the supermassive black hole at the Galactic Center, has long been examined in the radio-to-mm, near-infrared (NIR), and X-ray regimes of the electromagnetic spectrum. However, until now, sensitivity and angular resolution have been insufficient in the crucial mid-infrared (MIR) regime.
The MIRI instrument on JWST has changed that, and we report the first MIR detection of \Sg. The detection was during a flare that lasted about 40 minutes, a duration similar to NIR and X-ray flares, and the source's  spectral index steepened as the flare ended.
The steepening suggests that synchrotron cooling is an important process for \Sg's variability and implies magnetic fields strengths $\sim$40--70~Gauss in the emission zone.
Observations at $1.3~\mathrm{mm}$ with the Submillimeter Array revealed a counterpart flare lagging the MIR flare by $\approx$10~minutes.
The observations can be self-consistently explained as synchrotron radiation from a single population of gradually cooling high-energy electrons accelerated through (a combination of) magnetic reconnection and/or magnetized turbulence.
\end{abstract}

\section{Introduction}

\Sg\ is the source of electromagnetic radiation associated with the Milky Way Galaxy's central supermassive black hole. Of particular interest is its near-infrared and X-ray emission, which show variable emission with sporadic bright peaks in the light curve, phenomenologically called ``flares'' \citep{Baganoff2001_flare,Genzel2003,Ghez2004}.
Despite decades of research, the mechanism(s) behind these flares and their connection to \Sg's less-variable radio emission is not fully understood. Generally accepted to be caused by some energetic electron-acceleration event, plausible scenarios range from fluctuations in accretion rate and turbulence heating to magnetic reconnection \citep{Yuan2004, Yuan2009, Ball2016, Dexter2020_flare, Ripperda2020}.

A crucial missing observational link has been observations of flares in the mid-infrared (MIR), which has been the major gap between the near-infrared (NIR) and (sub-)millimeter regimes \citep{Schoedel2011}. The circumnuclear disk associated with \Sg\ was originally studied with ground-based MIR observations in the 1970's, especially in the 12.8~\micron\ [\ion{Ne}{2}] line \citep[e.g.,][]{Wollman1976}. 
\cite{Lacy1980} modeled the [\ion{Ne}{2}] line shapes and argued for the presence of a ${\sim}3 \times10^6$~\Msol\ point mass. 
However, these first MIR observations had limited the available spectral windows and lacked the high angular resolution and sensitivity now available with the James Webb Space Telescope (JWST). More recently, observations with the VISIR instrument \citep{Lagage2004_visir} at the Paranal observatory have provided stringent constraint on the flux density of Sgr A* at $8.6~\mathrm{\mu m}$ using a filter centered on the Paschen $\alpha$ line \citep{Schodel2011,Haubois2012,Dinh2024}. These ground-based observations, however, lacked the temporal stability to detect \Sg's variable flux in differential light curves and provide upper limits $20-50~\mathrm{mJy}$, depending on the choice of extinction correction.

The temporal evolution, and possible flux dependence of \Sg\ NIR spectral index, has been subject of intensive research. Evidence points towards a canonical index of $\approx + 0.5$ ($\nu L_{\nu}$) during bright flares \citep{Hornstein2007, DoddsEden2010, Ponti2017, GravityCollaboration2021_xrayflare}, with tentative evidence for negative indices at lower flux states and low level spectral index variations \citep{2006ApJ...640L.163G,2014IAUS..303..274W, 2018ApJ...863...15W}.

\section{Observations and Data Reduction}
\subsection{JWST MIRI/MRS}
The JWST observations were obtained on 2024 April 6 UT with the MIRI Medium Resolution Spectrometer (MRS) as part of Cycle~2 (Program ID 4572, PI D.\ Haggard merged with Program ID 3324, co-PIs: J.\ Hora, D.\ Haggard, G.\ Witzel). 
The MRS observes in four non-contiguous wavelength bands, in this case covering a spectral range 4.9 to 20.1~\micron\ in four bands separated by three 
gaps.\footnote{\url{https://jwst-docs.stsci.edu/jwst-mid-infrared-instrument/miri-observing-modes/miri-medium-resolution-spectroscopy\#gsc.tab=0}} 
The observations used the integral field unit, which observes a 3\farcs2$\times$3\farcs7 field of view (FoV) in Channel~1 (the shortest-wavelength band) and successively larger FoVs in the longer-wavelength channels.  Pixel sizes are 0\farcs196 in Channels~1 and~2, 0\farcs245 in Channel~3, and 0\farcs273 in Channel~4. The spectral resolution is $\sim$3500--1700, but for present purposes, the full spectrum in each channel was averaged to a single value representing flux densities at 5.3, 8.1, 12.5, and 19.3~\micron.

\begin{figure}
    \centering
    \includegraphics[width=0.47\textwidth]{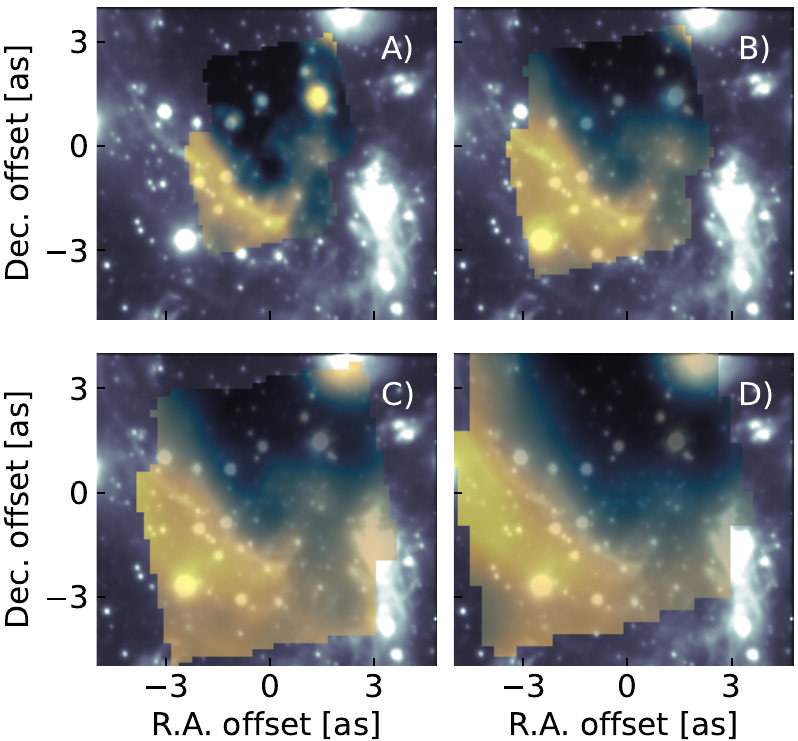}
    \caption{Mid infrared images of the Galactic Center with JWST. The color insets show JWST $5.3~\mathrm{\mu m}$ (A), $8.1~\mathrm{\mu m}$ (B), $12.5~\mathrm{\mu m}$ (C), and $19.3~\mathrm{\mu m}$ (D) observations. 
    The JWST/MIRI images are superposed on the $L'$ (NIR 3.8~\micron) stellar-background image from the Keck Observatory \citep{Hora2014}.
    The image scale is labeled in arcseconds with \Sg\ at the origin. North is up, east to the left.}
    \label{fig:observation_overview}
\end{figure}

This paper analyzed a part of the light curve from 10:39:35 UT to 12:37:09 UT\null. 
Time at the solar-system barycenter was 243~seconds later.\footnote{Keyword \textsc{bartdelt} in the processed Level 3 file header.}
The MRS data used here were taken in a single exposure of 85 integrations. This exposure was divided into five segments, the first four consisting of 18 integrations and the last one having 13 integrations.
We calibrated the MIRI data using the default JWST pipeline \citep{jwst_pipeline_bushouse2024}  version 1.16.0 in context of pmap jwst-1242 starting with the Level~2 data products downloaded from the MAST archive. To use the integration-level data (86.04-second cadence), we exported each level~2 segment into its individual integrations. 
The integrations were converted into 85 flux-calibrated 3D data cubes using the pipeline routine \textsc{calwebb\_spec3}, which uses the drizzle algorithm \citep{Smith2007, Law2023} to create an optimum data cube (R.A./Decl./wavelength) from the 2D image on the MRS detector. In each of the 85 data cubes, we masked wavelengths affected by emission lines, and the remaining flux densities in each of the four MRS wavelength bands were averaged. This gave light curves at four wavelengths with 85 data points each. Images in the four MRS channels are shown in Figure~\ref{fig:observation_overview}.

The calibration was refined by computing reference light curves from all pixels that showed a median flux level within $10\%$ of the median flux in the \Sg\ pixel (within $15\%$ for channel~4, where fewer than 10 pixels were within $10\%$) and $\geq$3 pixels away from the \Sg\ pixel so as to remain outside its PSF\null. A detailed discussion of the reduction and lightcurve extraction are given in Appendix \ref{apx:MIRI}.

The data show characteristics of the known systematics of the MIRI/MRS system,\footnote{\url{https://jwst-docs.stsci.edu/known-issues-with-jwst-data/miri-known-issues/miri-mrs-known-issues\#gsc.tab=0}} including fringing in the spectra as well as a quasi-periodic modulation in the spectra due to the MRS undersampling of the point-spread function. Both effects introduce errors in the observed flux density as a function of wavelength, but these are constant and therefore cancel in the normalized light curves. However, both effects can affect the light curve if the pupil illumination changes, i.e., if pointing errors occur. During the observed flare, the pointing was stable, and the effect was negligible. Nevertheless, we mitigated the effect of fringing by using the pipeline routine \texttt{fit\_residual\_fringes\_1d}. The recommended strategy for mitigating the PSF undersampling is to use a larger flux extraction aperture. However, in our case, sampling more than one pixel increased the noise in the reference light curves (\citealt{Hora2014} found the same result in IRAC data). Therefore, we derived the \Sg\ light curves from a single pixel. 
To convert the pixel value of surface brightness in MJy/sr to the total flux of a point source at that location in mJy, we multiply the pixel value by the encircled energy ratio assuming a Gaussian point spread functions for each channel given in the JWST MIRI documentation ($f_{\rm{ch1}}=2.74\times10^{-3}$; $f_{\rm{ch2}}=4.40\times10^{-3}$; $f_{\rm{ch3}}=7.66\times10^{-3}$; $f_{\rm{ch4}}=19.84\times10^{-3}$).

\begin{figure*}
    \centering
    \includegraphics[width=0.8\textwidth]{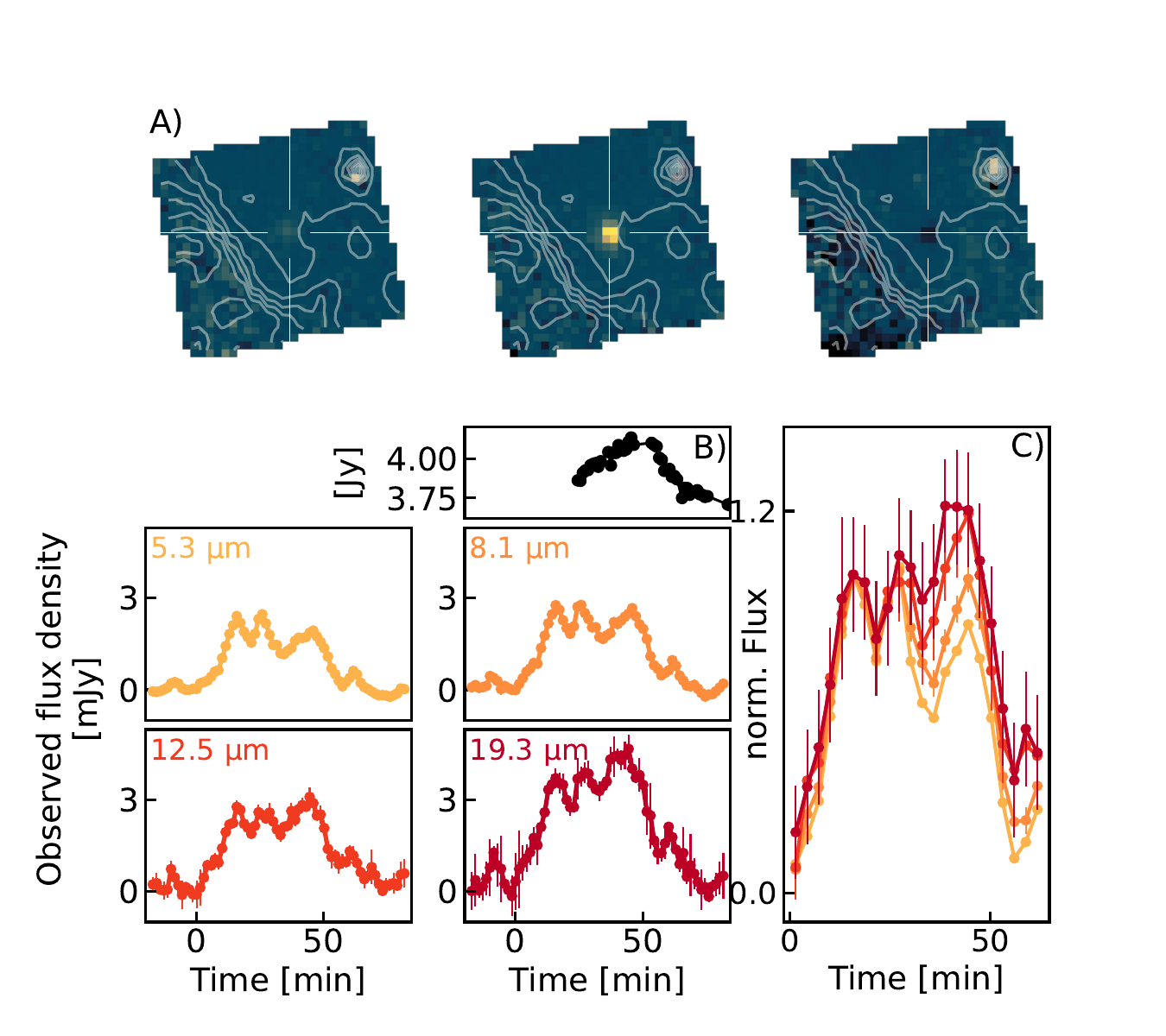}
    \caption{\textbf{Images and light curves of \Sg\ observed on 2024-04-06.} Panel A (top) shows  $8.1~\mathrm{\mu m}$ residual images after subtracting a constant image shown by the grey contour lines. Each image shows ${\sim}20~\mathrm{minutes}$ of averaged data before (left), during (middle), and after (right) the flare. North is up, and east is to the left. Panels B show light curves of \Sg\ at 1.3~mm (top left and right duplicated) and the four MIRI spectral channels with wavelengths as labeled.  Flux densities are as observed, i.e., calibrated but with no extinction correction. $t=0$ marks the beginning of the flare as we define it, and all times have been corrected to the barycentric reference frame. Panel C shows normalized light curves of the four MIRI bands to demonstrate the change in spectral index.}
    \label{fig1:LC}
\end{figure*}

We estimate a False Alarm Rate (FAR) for detecting a (stochastic) change in spectral index with comparable significance in our reference light curves by computing $10{,}000$ bootstrapped sets of CH1 to Ch4 reference light curves. We fit the spectral index for each set for all light curve points. We create $190{,}000$ such spectral index measurements. 
We find $2418$ measurements significantly different from zero at $1\sigma$,  $2$ at $2\sigma$. Since our signal is significant at $<4\sigma$, the FAR  $<190{,}000^{-1}$.

\subsection{SMA 220~GHz Observations}
The SMA observation of \Sg\ (project code 2023B-S017, P.I. H.\ Smith) began on-source at 11:44~UT and ended at 15:43~UT\null. These times were corrected to TDB using astropy times and EarthLocation sites, which gave a correction  of
$+$172~seconds. The SMA was in the extended configuration with a central tuning of 220.1 GHz, split into upper and lower sidebands consisting of six contiguous spectral windows. Each spectral window had 2.288~GHz bandwidth in 4096 spectral channels. We manually calibrated the total intensity (Stokes~I) data in CASA (version 5.6.2-3, \citealt{McMullin2007}) following standard interferometric calibration. Before calibration, we flagged channels showing RFI spikes on each baseline and kept only the inner 95\% of channels in each spectral window. 3C~279 was used as the bandpass calibrator, and Ceres set the absolute flux level. We used the JPL/Horizons model of Ceres \citep{Butler2012} and transferred the solutions to the three gain calibrators: J1751+096, J1733$-$130, and J1924$-$292. We solved for scan-averaged gain solutions on these calibrators and applied them to \Sg.

We split the calibrated \Sg\ visibilities for point-source self-calibration on all baselines and exported the self-calibrated data and the gain calibrator visibilities into AIPS for light-curve extraction. We used the AIPS task \textsc{dftpl} to extract light curves in the upper and lower sidebands at 60-second binning on projected baselines $\geq$30~k$\lambda$ to suppress any potential contamination by the surrounding extended emission.  To obtain a realistic estimate of the measurement uncertainty, we calculated the RMS of all calibrator light curves.

\subsection{Chandra X-ray Observations}
The Chandra observation of \Sg\ (Program 25700310, PI D.\ Haggard, ObsID 28230) was taken in FAINT mode using a 1/8 subarray on the ACIS-S3 chip for a total exposure time of 25.09~ks. The small subarray and shorter frame rate decrease photon pileup during flares.
The Chandra observation times were barycenter-corrected using the \textsc{axbary} method from the \textsc{CIAO v4.16} software package. The script calculates the correction using the ICRS reference system and applies the value to the TIME column in the observation's event file. This was performed immediately after recalibration so adjustments were carried downstream and applied to the resulting analysis such as lightcurve extraction. In this observation, the median Chandra barycenter correction was $+$172~seconds.

\begin{figure*}
    \centering
    \includegraphics[width=0.95\textwidth]
    {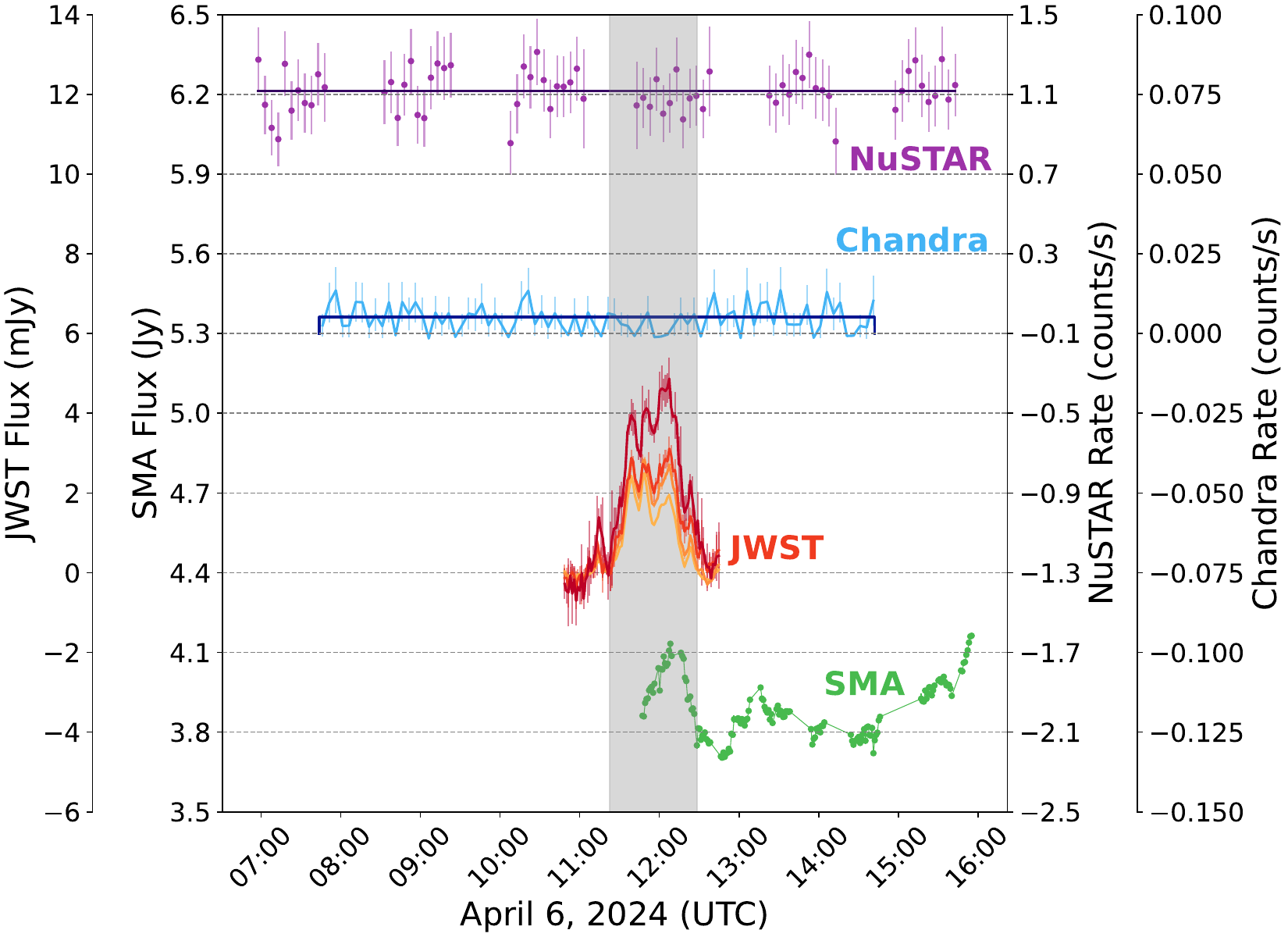}
    \caption{\textbf{Full multi-wavelength \Sg\ light curves on 2024 April 6.} From bottom to top: 
    the green curve shows  
    the average of the SMA upper (230~GHz) and lower (210~GHz) sidebands. 
    The superposed yellow to red curves show the observed  flux densities in JWST/MIRI channels 1 to 4, respectively. The blue curve shows the Chandra 2--8~keV light curve (corrected for background and pileup) with 300~s binning, and the horizontal dark blue line shows the median of the entire Chandra observation. 
    The purple curve shows the NuSTAR 3--30~keV light curve with 300~s binning. The dark purple overlay is the average during the observation period.
    Each light curve has its own ordinate as labeled, and times are UTC at the Solar System barycenter.
    No X-ray flares were detected by either Chandra or NuSTAR. The grey region highlights the MIR flare interval seen by JWST.}
    \label{fig:MWL_LCs}
\end{figure*}

Chandra data reduction followed \cite{Boyce2022} and used the \textsc{CIAO v4.16} software package. We first reprocessed the level~2 event file with the latest calibration using the \textsc{chandra\_repro} script (CALDB v4.11.2) and then applied the barycentric correction. We extracted a 300~s-binned 2--8~keV background-subtracted light curve within a circular 1\farcs25 radius region centered on \Sg\ and a circular background annulus of inner radius 14\arcsec\ and outer radius 20\arcsec\ (Figure \ref{fig:MWL_LCs}). The small extraction region decreases background events and X-ray emission from nearby sources. 
While severe pileup was mitigated by the observing setup, we applied the pileup correction method of \cite{Bouffard2019} to compensate for any remaining effects. 
We searched for X-ray flares using the Bayesian Blocks algorithm \citep{Scargle2013} but did not observe any flares in the Chandra exposure. \Sg's average X-ray flux during the entire observation was 0.005 counts~s$^{-1}$ corresponding to $2.9\times10^{-13}$~ergs cm$^{-2}$ s$^{-1}$ (Figure~\ref{fig:MWL_LCs}).

\subsection{NuSTAR X-ray Observations}
NuSTAR X-ray observations on 2024 April 6 were collected as part of the \Sg\ multi-wavelength campaign ((Program 9041, PI S.\ Zhang, ObsID 30902013004). We reduced the NuSTAR data using \textsc{NuSTAR-DAS} (v.2.1.1) and \textsc{HEASOFT} (v.6.32) to extract 3--30~keV barycenter-corrected (${\sim+}172$~seconds) light curves from a 30\arcsec\ region at the position of \Sg\ with 300~s binning (Figure~\ref{fig:MWL_LCs}).  
Following \cite{Zhang2017}, we also ran a Bayesian Blocks search for flares in the NuSTAR lightcurves but did not detect any X-ray flares. 
The combined Focal Plane Module (FPMA$+$FPMB) count rate from the source region in 3--30~keV was 1.1 counts s$^{-1}$ corresponding to 
an observed flux of $9.6\times10^{-12}$ ergs cm$^{-2}$~s$^{-1}$. The emission is background dominated, with \Sg\ contributing $\sim$5\% of the total signal.

\section{Results}
The MIR light curves (Figure~\ref{fig1:LC}) revealed a bright flare seen in all four bands and lasting about 50~minutes, similar to the duration of NIR flares  \citep[e.g.,][]{vonFellenberg2023_sgra, vonFellenberg2024}. 
No variability was observed at X-ray wavelengths. NIR flares without X-ray emission are common \citep[e.g.,][]{Boyce2019}, and it is not surprising to see the same occurrence for a MIR flare.
The SMA began observing \Sg\ about 10-minutes into the MIR flare and saw the 1.3~mm flux density increase by $\sim$0.3~{Jy} compared to its initial value ((Figure~\ref{fig1:LC} and~S4).  The discussion below argues that the mm-wave increase was most likely connected to the MIR flare. 

Due to the bright, surrounding thermal dust MIR emission in the Galactic Center, \Sg\ in its quiescent state is not detectable; only its variable emission is detected in the difference images (Figure~\ref{fig1:LC}). In addition, because the strong \citep[about 3~magnitudes at 10~\micron; e.g.,][]{Fritz2011} dust extinction varies both spectrally and spatially, the MIR extinction to \Sg\ is uncertain,  complicating the derivation of an absolute flux-calibrated spectrum of \Sg\ and making it prone to systematic biases. To mitigate this problem, we adopted a spectro-differential approach: we normalized the four MIR light curves between their median flux levels and the bright flux levels at one moment near the beginning of the flare. The resulting normalized spectral energy distribution (SED) is independent of the wavelength-dependent extinction and the unknown static flux subtracted from each light curve. 
While this prevents absolute spectral-index measurements, it is a robust way to measure the relative light curves and the time evolution of the spectral index.

The MIR flare can be divided into three phases (Figure~\ref{fig:flare_fit}): 1) a fast rise (${\approx} 10~\mathrm{minutes}$), 2) a falling-and-rising phase \textbf{(${\approx} 20~\mathrm{minutes}$)}, and 3) a fast decay of the flux density at the end (${\approx} 10~\mathrm{minutes}$).
The light curves of the flare in MIRI's four bands were not the same (Figure~\ref{fig:spectral index}), and \Sg's
spectral index $\alpha$ (defined by $F_{\nu} (t)\propto \nu^{\alpha(t_0) + \Delta \alpha (t)}$) changed systematically during the flare. 
During the first phase, there was no measurable change in spectral index, $\Delta \alpha \approx 0$. During the second phase, the spectrum reddened by $\Delta \alpha\approx -0.4\pm0.1$. This slope remained similar until the flare ended. This is the first measurement of a significant change in spectral index during a bright phase of a \Sg\ flare. 

\section{Flare Modeling}
\subsection{Description of Model}
A simple flare model can explain the temporal evolution of \Sg's spectral index. Details of the model are given in Appendix \ref{Apx:models}. Our model assumes a particle acceleration event where electrons are continuously injected into an emission region of constant size with the injection rate following a Gaussian profile. 
The magnetic field in the emission region is assumed constant.
The injected electrons have a power-law energy distribution $dN/dE \propto \gamma^{-p}$ for $\gamma \in [\gamma_{\rm{min}}, \gamma_{\rm{max}}]$, where $dN$ is the number density of electrons in an energy range $dE$, $\gamma$ is the electron Lorentz factor, and $p$ is a free parameter.
During and after the injection event, the electron energy distribution evolves by electron-synchrotron cooling \citep[e.g.,][]{Dodds-Eden2010}, where the synchrotron cooling time scale is given by:
\begin{equation}
    t_{\rm{sync}} (B, \nu) \approx 8 \left( \dfrac{B}{30~\rm{G}} \right)^{-3/2} \left( \dfrac{\nu}{10^{14}~\rm{Hz}} \right)^{-1/2} \mathrm{minutes}\quad,\label{eq:sync_cool}
\end{equation}
for a fiducial magnetic field strength $B=30$~G.
%
The model includes Doppler boosting and de-boosting from a circular orbit of the emission region around the central black hole as well as a gravitational-redshift term. The  amplitude of these effects depends on the observer angle $\phi_{\rm{inc}}$, the radial separation of the emitting region from the black hole $R_{\rm{orb}}$, and the initial angular position of the emitting region $\Omega_0$. Gravitational lensing was ignored in our model, a valid assumption if the orbit is viewed at inclination $\phi_{\rm{inc}} \lesssim 50^{\circ}$ and the motion is mostly in the plane of the accretion flow \citep[e.g.,][]{GravityCollaboration2020_orbital}. The inclusion of Doppler boosting is motivated by the GRAVITY observation of low-inclination orbital motion \citep{GRAVITYCollaboration2018_orbital, GravityCollaboration2020_orbital, GRAVITY_collab2023_polariflares}, the EHT image of the black-hole shadow \citep{eht_sgra_I, eht_sgra_II, eht_paper_III, eht_paper_IV, eht_paper_V, eht_paper_VI}, and the mm observations in the context of the EHT observations \citep{Wielgus2022}, all of which suggest that Doppler boosting must be an important contributor to the overall variability \citep{vonFellenberg2023_sgra, vonFellenberg2024}. 

According to the model, in the first phase of the flare the fast, achromatic rise in flux is caused by the growing number of injected electrons together with Doppler boosting. 
In the second phase, the reddening of the spectral index by $\Delta \alpha \approx -0.4$ arises from synchrotron cooling of high-energy electrons to lower energies. The decrease and subsequent increase in flux is explained by Doppler de-boosting first and then boosting that occurs one orbit after the initial injection. 
In the third phase, the initial decline comes from efficient cooling, and the decline speeds up as Doppler de-boosting sets in. 
The change in spectral index and the fast decay at the end of the flare constrain the total duration of the electron acceleration event. While we do not know the intrinsic injection profile, a Gaussian profile fits the observed flare shape when allowing for Doppler boosting due to orbital motion. The best-fit injection timescale has a Gaussian root-variance width $\sigma = 11.7_{-0.2}^{+0.2}~\mathrm{minutes}$. 

Doppler deboosting alone cannot explain the flare’s fast decay. Assuming an orbiting hot-spot model, where there is no net outflow and/or expansion of the emission region, the flare’s fast decay requires the removal of emitting electrons either by particle escape or by efficient electron cooling through the emission of synchrotron radiation. 
Particle escape is unlikely because MIR-emitting electrons have a gyroradius smaller than the Schwarzschild radius and, therefore, should be confined in the turbulent emission region \citep{Lemoine2023,Kempksi2023}. 
However, particles can cool efficiently on timescales of $\approx$5~minutes within the emitting region provided the magnetic field is high, $B \approx 40$--70~G (Equation~\ref{eq:sync_cool}) for MIR-emitting ($\nu_{\rm{JWST}} \approx (\hbox{1.5--6}) \times10^{13}~\mathrm{Hz}$) electrons.

Depending on the underlying model assumptions, the best fit magnetic field strength ($B \approx 40$--70~{G}) required for synchrotron cooling in the flare emission zone is slightly higher than the $\sim$30~G typically suggested \citep[e.g.,][]{Dodds-Eden2010,Ponti2017,GravityCollaboration2021_xrayflare}. However, the higher field strength is consistent with those earlier modeling results. Previous ground and space observations lacked the sensitivity and spectral range to measure the spectral evolution during flares, and absent spectral information, the magnetic field strength is largely degenerate with the size of the emission region and the particle number density within it \citep[e.g.,][]{GravityCollaboration2021_xrayflare}.

\begin{figure}
    \centering
    \includegraphics[width=0.47\textwidth]{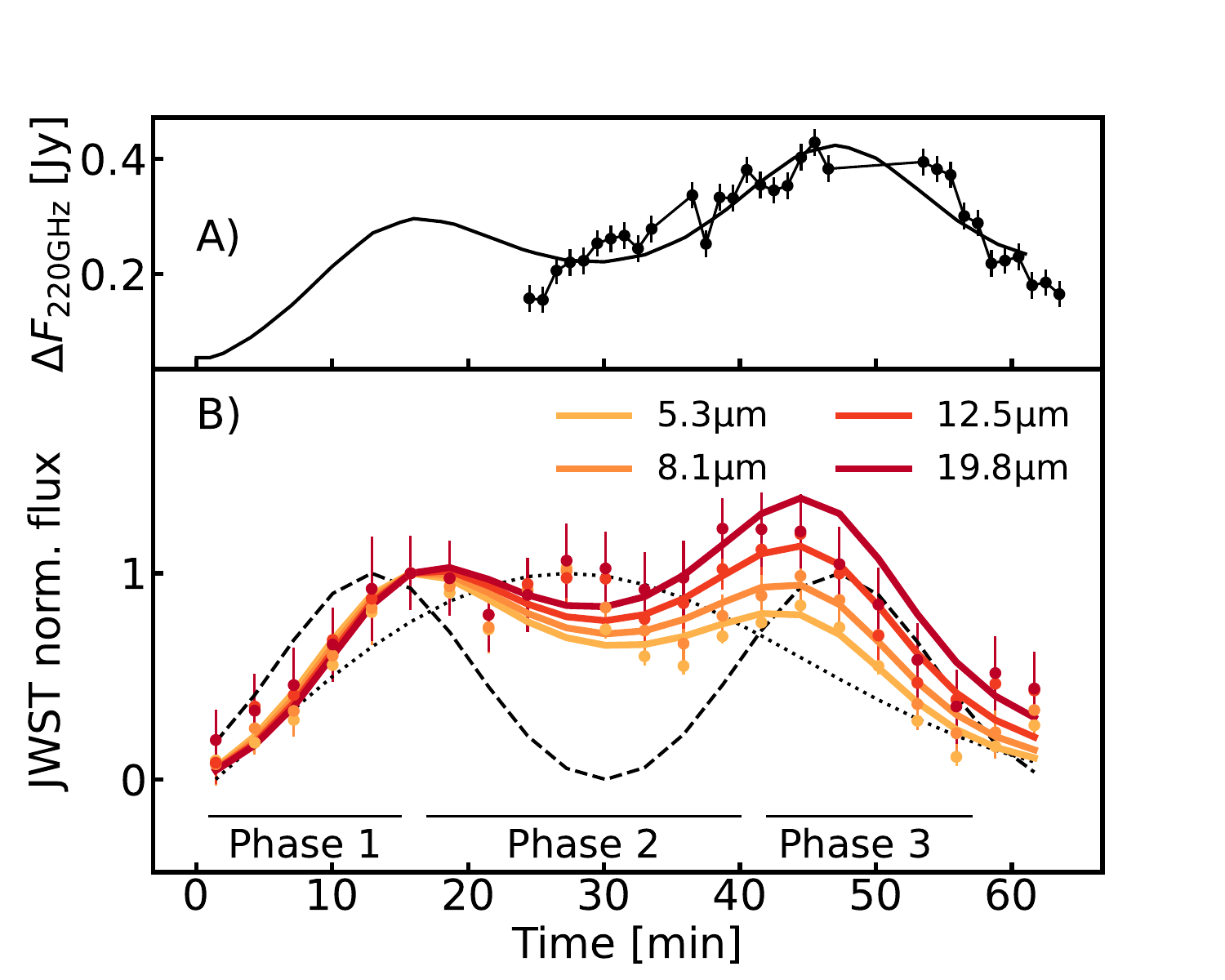}
    \caption{\textbf{Observed light curves with model fit.} Points show the observed data, and solid lines show the predictions of the simple model described in the text. (A)  220~GHz data.
    (B) MIR data. The four yellow-to-red lines show the best-fit model predictions for each channel as labeled. The black dashed line shows the modeled Doppler boosting and de-boosting due to the orbital motion of the emission region. The dotted line shows the modeled Gaussian injection curve of power-law electrons.  The flare phases described in the text are indicated at bottom.}
    \label{fig:flare_fit}
\end{figure}

\begin{figure}
    \centering
    \includegraphics[width=0.47\textwidth]{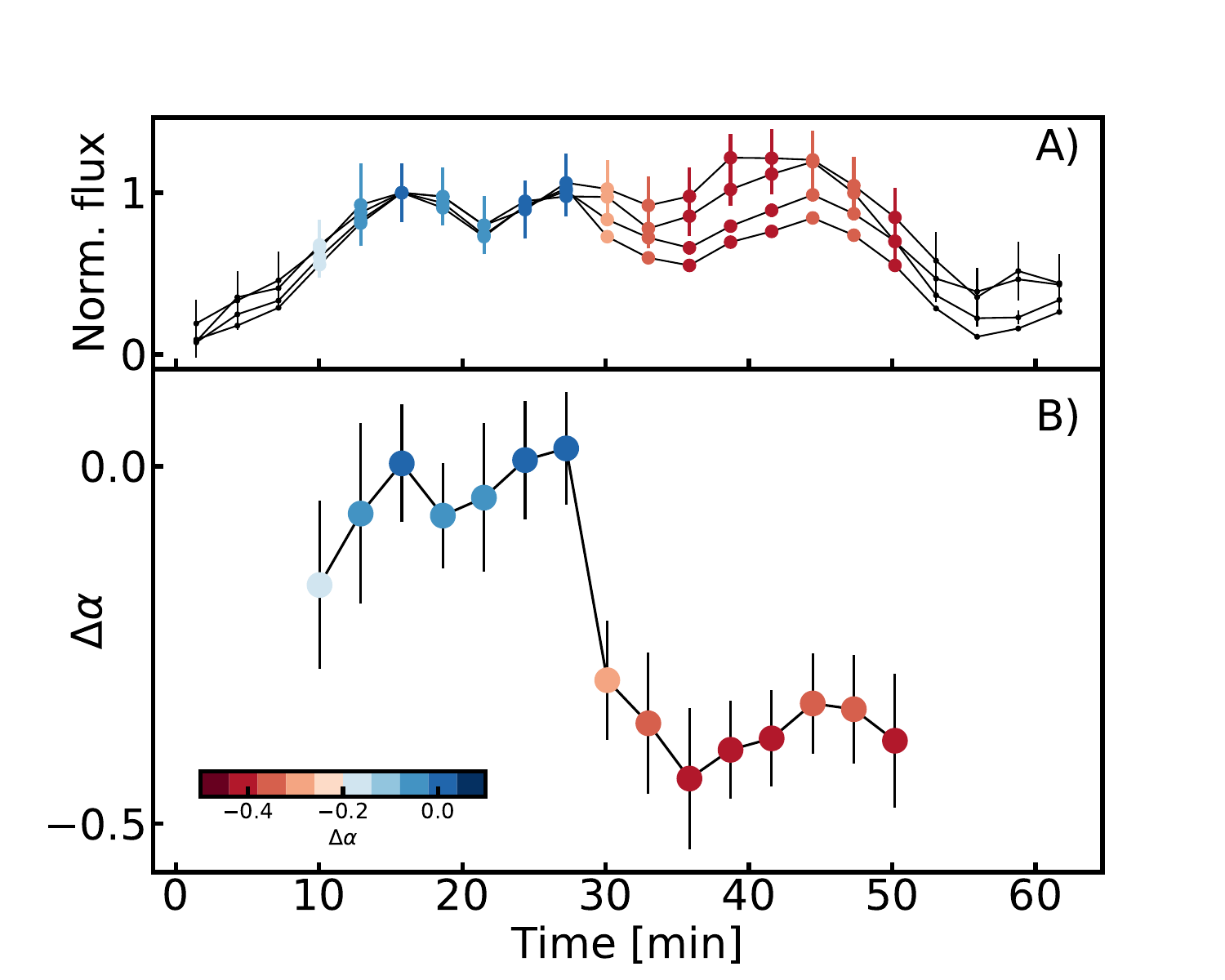}
    \caption{\textbf{Evolution of the spectral index during the flare.} Panel A) shows the normalized MIR light curves, and panel B) shows the change in the best-fit spectral index. The colors indicate the spectral index with blue colors $\Delta \alpha\approx 0.0$, and red colors $\Delta \alpha\approx -0.4$.}
    \label{fig:spectral index}
\end{figure}

A strong prediction of the model is that as the electrons cool to lower energies, they must emit synchrotron radiation at lower frequencies.
If $\gamma_{\rm{min}} =10$ and the total electron number density in the emitting region $n_{\e} \approx 1$--$10 \times 10^6 ~\mathrm{cm^{-3}}$, the model produces mm flux in the range of the observed values while at the same time fitting the observed MIR flare (Figure~\ref{fig:flare_fit}).
This result is a lower limit for the total number density, consistent with canonically quoted accretion-flow densities $\sim$10$^{6}$--10$^7~\mathrm{cm^{-3}}$ \citep[e.g.,][]{Gillessen2019}.
Thus, synchrotron cooling is a plausible mechanism to explain the observed $\Delta F_{\rm{1.3~mm}}\sim 0.3~\mathrm{Jy}$ increase in mm flux density. 
However, the MIR data alone do not require a low $\gamma_{\rm min}$. If $\gamma_{\rm min}\gtrsim 10$, then the total mm flux is too low. The model also does not inherently require the electron numbers to be that high. If $n_{\e} < 1 \times 10^5~\mathrm{cm^{-3}}$,  the model's mm flux density would be $<$0.05~{Jy}, too small to be discernible in the mm light curve. Such models fit the MIR light curve, but the observed mm variability would have to be explained by an independent process, and the observed mm rise would be just a coincidence. 
Because \Sg's mm emission is constantly variable \citep[e.g.,][]{eht_paper_V}, with observed flux densities ranging from $\sim$2--6~Jy, a by-chance association is tenable, and the observed variability amplitude of $0.3~\mathrm{Jy}$ is consistent with the measured variability amplitude at 230 GHz \citep[e.g.,][]{Dexter2014}.
Nevertheless, the model's ability to produce detectable mm flux, assuming canonical accretion-flow parameters, favors a causal connection between the MIR and mm-flux density.

While the simple one-zone model has no specific scenario of electron acceleration, it reflects several aspects of electron acceleration in turbulence in a magnetized region in the accretion flow, e.g., in flux tubes in accretion disks \citep{Porth2021, Ripperda2022,Zhdankin2023,Grigorian2024}. 
In such a model, an eruption of magnetic flux from the black hole's event horizon produces an orbiting low-density cavity in the accretion flow, where the typical proton magnetization is of order $\sigma_{\rm p} = B^2 / (4\pi n_\p  m_\p c^2)\sim0.1$--1 \citep[e.g.,][]{Dexter2020_flare,Porth2021, Ripperda2022,NajafiZiyazi2024,Grigorian2024}. Due to Rayleigh--Taylor instability, these cavities are quickly protruded by streams of plasma from the surrounding accretion flow. Subsequently, reconnection within the protrusions can efficiently accelerate electrons to Lorentz factors capable of producing IR synchrotron emission \citep{Zhdankin2023}. In contrast to sporadic magnetic reconnection events, the Rayleigh--Taylor instability can continuously drive turbulence in the cavity, energizing electrons for as long as the cavity exists.

The number density of electrons in the non-thermal component and the magnetic field strength set an upper limit on the electron magnetization $\sigma_{\rm e} = B^2 / (4\pi n_\e  m_\e c^2) \approx 40$ for an emitting region of size $R_{\rm{flare}}=1~\mathrm{R_\mathrm{S}}$. Because the amount of synchrotron flux produced by the flare depends on the size of the emission region and the particle density, changing the size changes the electron magnetization $\sigma_{\rm{e}}$. 
We derive the posterior distribution of $\sigma_{\rm{e}}$ assuming that the cavity sizes range from $R_{\rm{flare}} \in [0.8\mathrm{R_\mathrm{S}}, 2.3R_\mathrm{S}]$ (i.e., a flat $R_{\rm{flare}}$ prior), based on typical sizes of orbiting low-density magnetic flux tubes in GRMHD simulations \citep{Porth2021,Dexter2020_flare,Ripperda2022}.
The derived posterior distributions imply $\sigma_{\rm{e}} > 10$ with an inter-quartile range $\mathrm{IQR} \approx 140$ (25\%), $380$ (50\%), $920$ (75\%).
Accounting for the higher proton mass, the proton magnetization $\sigma_{\rm p} > 0.006$, $\mathrm{IQR}\approx 0.1~(25\%)$, $0.2$ (50\%), $0.5$ (75\%). Values of $\sigma_{\rm p} \gtrsim 0.1$ are consistent with the proton magnetization found in magnetized cavities in GRMHD simulations \citep{Ripperda2022}, and with magnetic reconnection in the turbulent cavity as an electron acceleration mechanism.
Within the cavity, we expect reconnection with a significant component of the magnetic field orthogonal to the reconnecting field (the so-called ``guide’' field). Under these conditions, reconnection is expected to quickly accelerate electrons to Lorentz factors $\gamma \sim \sigma_{\rm e}$ \citep[e.g.,][]{Comisso2024}. This population of electrons would be able to emit at infrared wavelengths \citep{Zhdankin2023}.

\subsection{Model Results}
The synchrotron model predicts flux densities at all wavelengths from mm to X-ray, though by construction it predicts zero X-ray emission. (If an X-ray flare had been observed, a higher $\gamma_{\rm max}$ would have predicted X-ray emission, but we have not explored whether this would fit historically observed X-ray flares.) At mm wavelengths, an extra model component is needed to account for synchrotron emission by low-energy, thermal electrons. We did not model this component but included a constant offset in the model fit. Because we set the lowest point in the mm-light curve to 0.1~{Jy}, this offset is $\approx$0.1~Jy.

The synchrotron model describes the following characteristics of the data:
\begin{enumerate}
    \item The rising phase of the flare, where injected electrons and Doppler boosting cause a rapid rise of MIR flux.
    \item The falling-and-rising phase in the middle, where the Doppler de-boosting decreases the observed flux of the still-increasing intrinsic emission.
    \item The spectral-index change during the falling-and-rising phase of the flare, caused by high-energy electron cooling.
    \item The rapid decay of the emission at the end of the flare,  explained by continued electron cooling along with the decrease in electron-injection rate.
\end{enumerate}

One thing the simplest model cannot explain is the fast double-peak clearly seen in bands 1, 2, and 3 at the beginning of the flare ($t\approx 10$--20~minutes). To avoid biasing the fit by this double peak feature, we increased the error bars for two data points at $t=25.5~\mathrm{minutes}$ and $t=28.5~\mathrm{minutes}$ by a factor of $3$ in channels~1 and~2. 
This fast double-peak could be explained by a more complex (double) injection profile or by variation in the magnetic-field strength. Section \ref{sec:doubleGauss} demonstrates the former by including a second Gaussian injection event to describe the fast double-peak at the beginning of the flare.
As noted above, when taking into account the theoretical uncertainty in electron temperature and magnetic to electron pressure ratio, the estimated magnetic field strength from the flaring region is consistent with the range of magnetic field strengths inferred based on EHT observations \citep[i.e.,][]{eht_paper_V}.

\section{Conclusions}
All in all, the new MIR observations suggest that \Sg's MIR emission comes from synchrotron emission by a cooling population of electrons.  The particle acceleration could come from a combination of magnetic reconnection and magnetized turbulence. 
In addition, our observations indicate that variable emission immediately following high-energy flares can be caused by non-thermal processes.
This underlines the importance of including non-thermal processes even for modeling lower-energy radio emissions. A complete physical model will need to include both global accretion-disk dynamics, which can provide particle acceleration sites, as well as microscopic energization processes and their role in powering the observed radiation \citep[e.g.,][]{Galishnikova2023,Zhdankin2023}.

\vskip 30pt

{We thank Charles Gammie and Michi Bauboeck for their comments and discussion on the manuscript.
This research was supported by the International Space Science Institute (ISSI) in Bern, through ISSI International Team project \#24-610, and we thank Mark Sargent and his team for their generous hospitality.
This work is based on observations made with the NASA/ESA/CSA James Webb Space Telescope. The data were obtained from the Mikulski Archive for Space Telescopes at the Space Telescope Science Institute, which is operated by the Association of Universities for Research in Astronomy, Inc., under NASA contract NAS 5-03127 for JWST. These observations are associated with program \#4572. The Submillimeter Array is a joint project between the Smithsonian Astrophysical Observatory and the Academia Sinica Institute of Astronomy and Astrophysics and is funded by the Smithsonian Institution and the Academia Sinica. We recognize that Maunakea is a culturally important site for the indigenous Hawaiian people; we are privileged to study the cosmos from its summit.}
This research has made use of data obtained from the Chandra Data Archive provided by the Chandra X-ray Center (CXC).

DH, RZS, NMF acknowledge support from the Canadian Space Agency (23JWGO2A01), the Natural Sciences and Engineering Research Council of Canada (NSERC) Discovery Grant program, the Canada Research Chairs (CRC) program, the Fondes de Recherche Nature et Technologies (FRQNT) Centre de recherche en astrophysique du Québec, and the Trottier Space Institute at McGill. NMF acknowledges funding from the FRQNT Doctoral Research Scholarship.

Support for program \#4572 was provided by NASA through a grant from the Space Telescope Science Institute, which is operated by the Association of Universities for Research in Astronomy, Inc., under NASA contract NAS 5-03127.

AP is supported by a grant from the Simons Foundation (MP-SCMPS-00001470). A.P. additionally acknowledges support by NASA grant 80NSSC22K1054.

BR, BSG are supported by the Natural Sciences \& Engineering Research Council of Canada (NSERC), the Canadian Space Agency (23JWGO2A01), and by a grant from the Simons Foundation (MP-SCMPS-00001470). BR acknowledges a guest researcher position at the Flatiron Institute, supported by the Simons Foundation.

SZ And GS acknowledge funding support from the NASA grant \#80NSSC23K1604. We thank the NuSTAR team for their efforts with the observation scheduling. This research has made use of software provided by the High Energy Astrophysics Science Archive Research Center (HEASARC), which is a service of the Astrophysics Science Division at NASA/GSFC.

JM is supported by an NSF Astronomy and Astrophysics Postdoctoral Fellowship under award AST-2401752. This research was supported in part through the computational resources and staff contributions provided for the Quest high-performance computing facility at Northwestern University, which is jointly supported by the Office of the Provost, the Office for Research, and Northwestern University Information Technology.

TR acknowledges funding support from the Deutscher Akademischer Austauschdienst (DAAD) Working Internships in Science and Engineering (WISE) program.

The observations are available at the Mikulski Archive for Space Telescopes (\url{https://mast.stsci.edu/}) under proposal IDs 4572 for JWST and at the Chandra Data Archive (\url{https://cxc.harvard.edu/cda/}) proposal number 25700310 for Chandra. The source code used for the flare model is available at \url{https://github.com/ydallilar/flaremodel}.


\bibliography{refs}{}
\bibliographystyle{aasjournal}
\clearpage

\appendix
\restartappendixnumbering
\section{MIRI/MRS Reduction Details}\label{apx:MIRI}
\begin{figure}[h]
    \centering
    \includegraphics[width=0.75\textwidth]{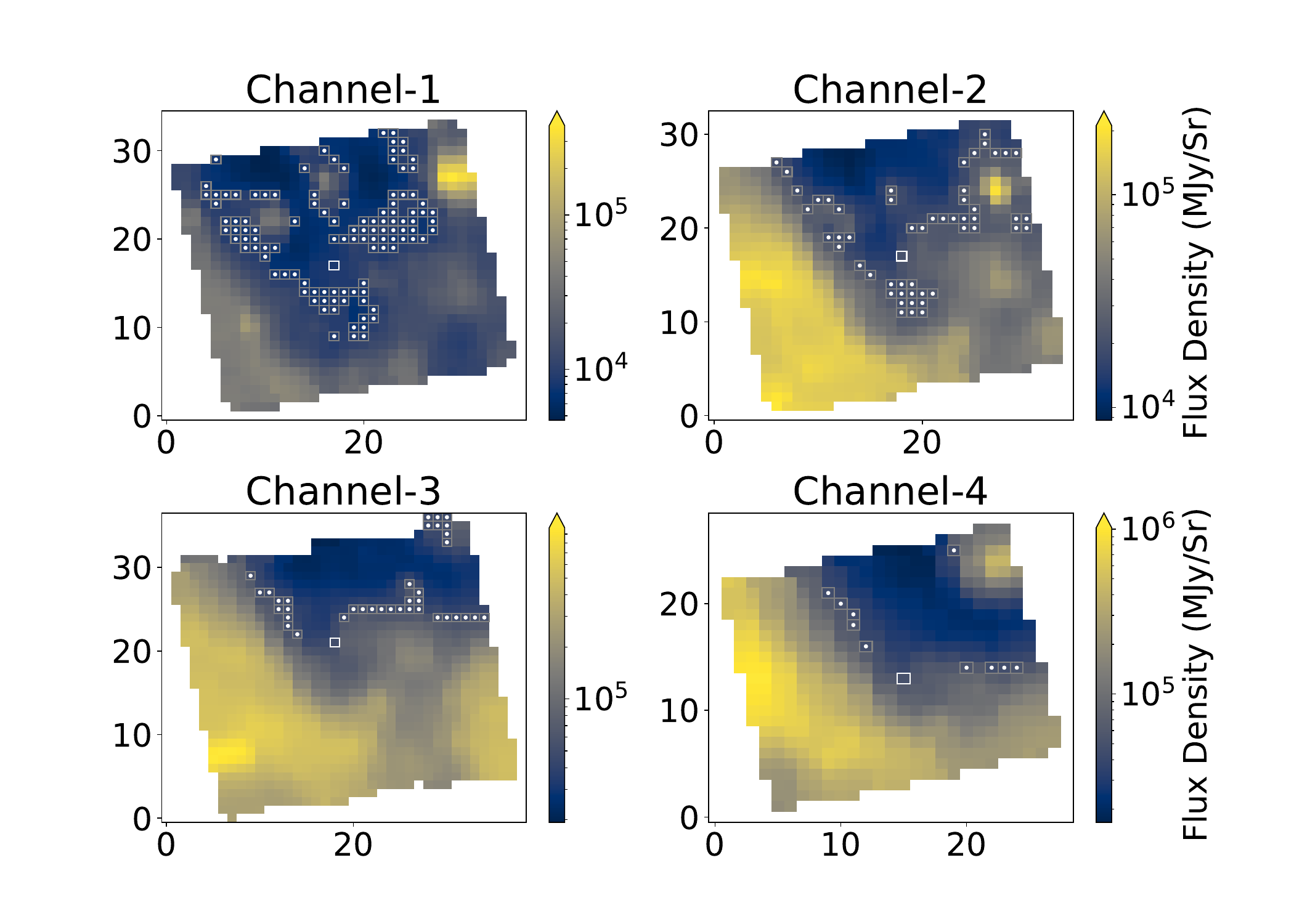}
    \caption{\textbf{Median images of the Galactic Center  in channels 1, 2, 3, and 4.} The location of \Sg\ is marked with the square box, and the locations of the reference pixels used to estimate the noise in the image are marked by white dots. Axis labels are in pixel number with pixel angular sizes given in the text. The flux-density scale is in MJy/sr.}
    \label{fig:reference_pixels}
\end{figure}

\begin{figure}[h]
    \centering
    \includegraphics[width=0.65\textwidth]{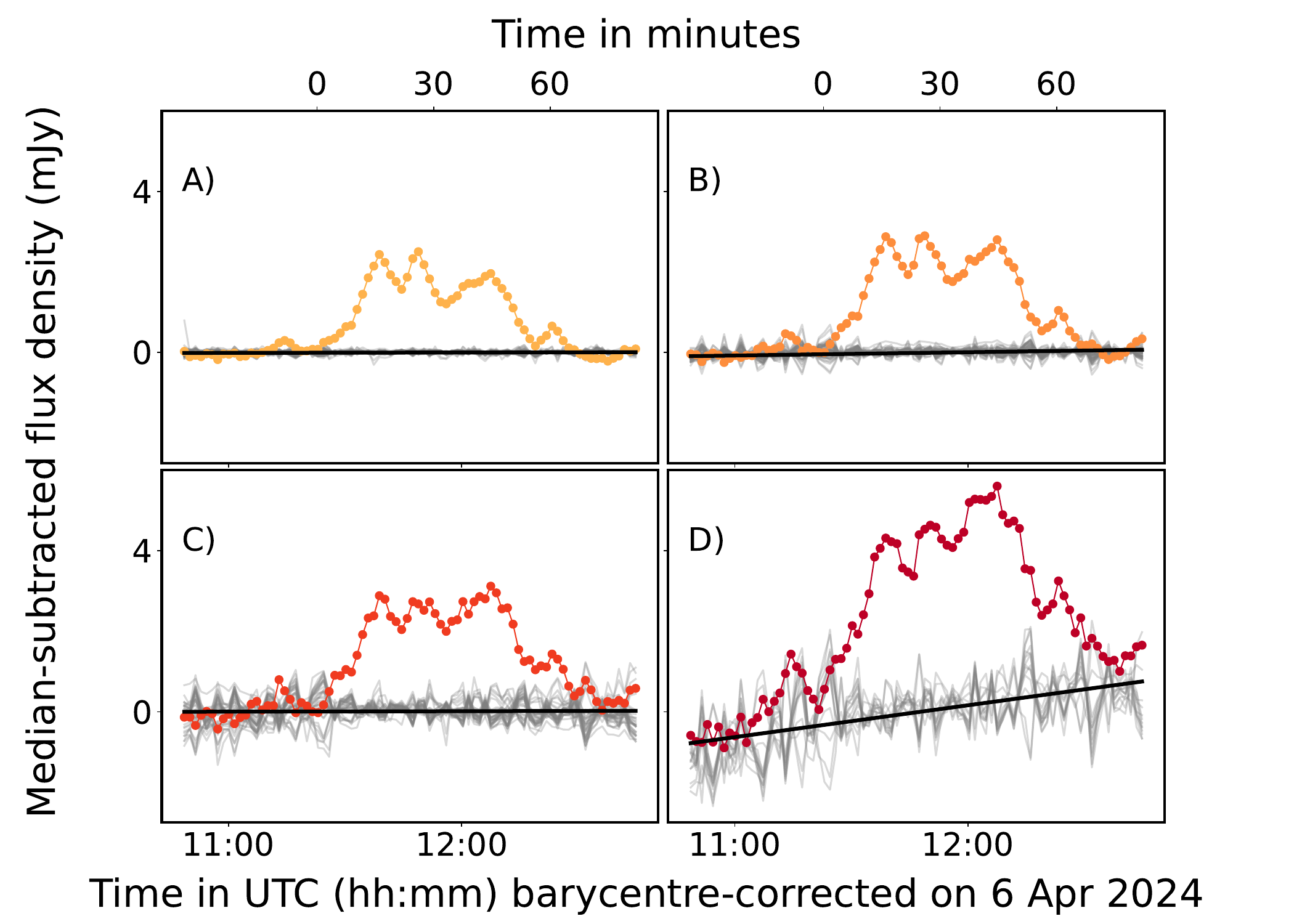}
    \caption{\textbf{Comparison of \Sg\ and reference-pixel light curves.} Panels show the median-subtracted light curves of the \Sg\ pixel (color) and reference pixels (grey) in channels 1, 2, 3, and 4 (panels A, B, C and D respectively). The black lines show the best linear fit to each reference-pixel trend. Table~\ref{tab:reduction} gives the line slopes. Times are shown in UTC at the Solar System barycenter on the lower abscissa and relative to the adopted $T=0$ on the upper abscissa.}
    \label{fig:lightcurve_processing}
\end{figure}

\begin{figure}[h]
    \centering
    \includegraphics[width=0.65\textwidth]{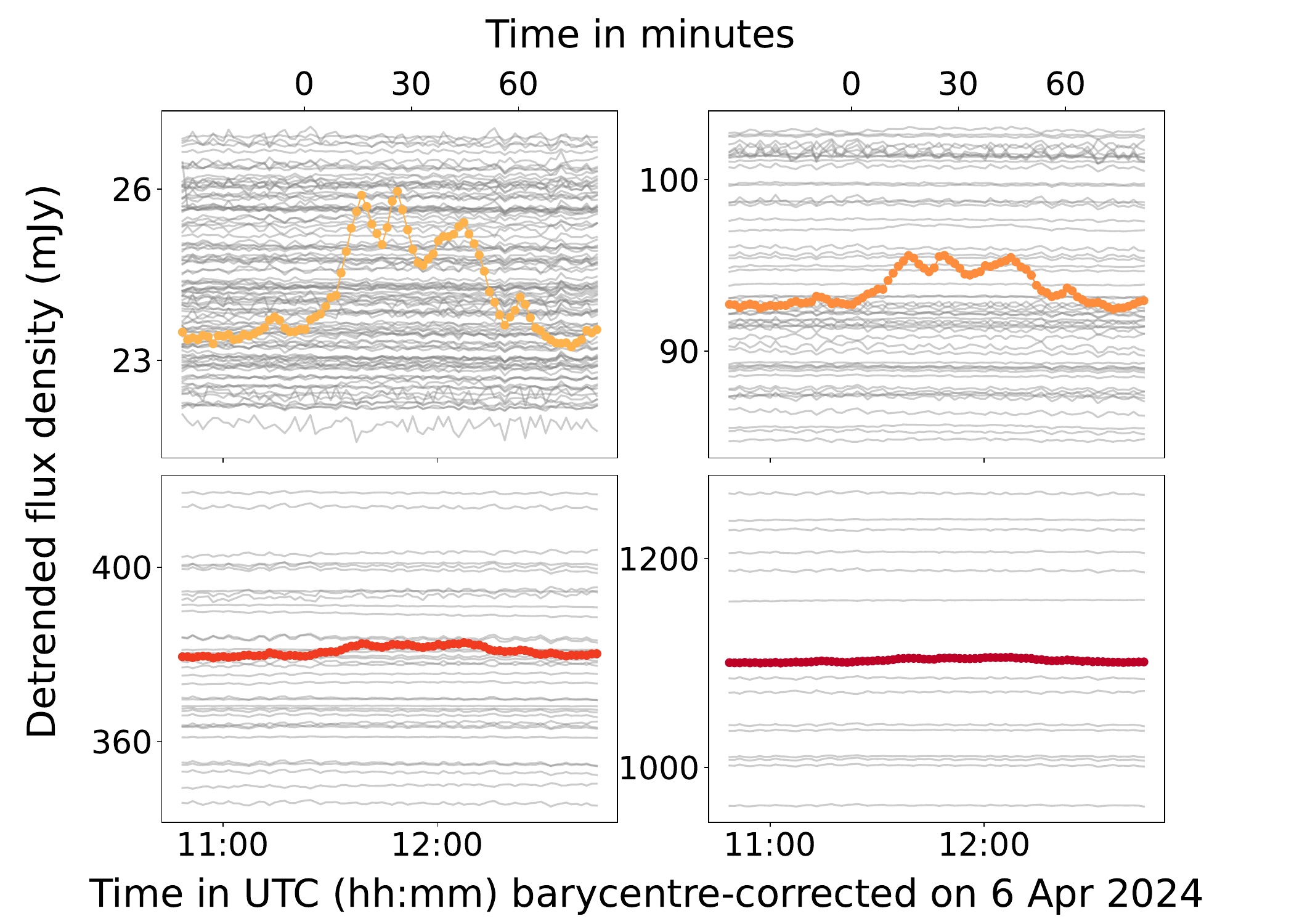}
    \caption{\textbf{Light curves of reference pixels compared with the \Sg\ pixel.} Grey curves in each panel (channels 1, 2, 3, 4 in panels A, B, C, D respectively) show the reference-pixel light curves after removing the linear slope but without subtracting each pixel's median value.  Colored lines show the same for the \Sg\ pixel.}
    \label{fig:reference_absolute}
\end{figure}

The calibration was refined by computing reference light curves from all pixels that showed a median flux level within $10\%$ of the median flux in the \Sg\ pixel (within $15\%$ for channel~4, where fewer than 10 pixels were within $10\%$) and $\geq$3 pixels away from the \Sg\ pixel so as to remain outside its PSF\null. 
Figure~\ref{fig:reference_pixels} shows the reference-pixel locations. The only systematic effect within the exposure was a linear drift of the pixel zeropoints.
We corrected that by fitting a linear function to each reference pixel's light curve, averaging the slopes, and subtracting the corresponding flux from the \Sg\ pixel flux of each point in the \Sg\ light curves. Figure~\ref{fig:lightcurve_processing} shows the de-trending result, and Table~\ref{tab:reduction} gives the measured slopes along with the RMS of the reference-pixel light curves and the flux offset subtracted from the \Sg\ data. Figure~\ref{fig:reference_absolute} shows the light curves of \Sg\ and the reference pixels after correction for the average slope but not each pixel's median. This shows that the linear drift does not depend on the absolute flux values, i.e., it was a zeropoint drift rather than a gain drift.
In order to estimate the systematic uncertainty introduced by the drift correction, we generated bootstrapped surrogates of the reference pixels and calculated the drift correction for each bootstrapped sample. We then calculated the RMS of the resulting light curves and took the maximum RMS value as the uncertainty reported in Table~\ref{tab:reduction}. Because these uncertainties are small with respect to the photometric uncertainty, we ignore this source of systematic uncertainty.

\begin{figure}[h]
    \centering
    \includegraphics[width=0.65\textwidth]{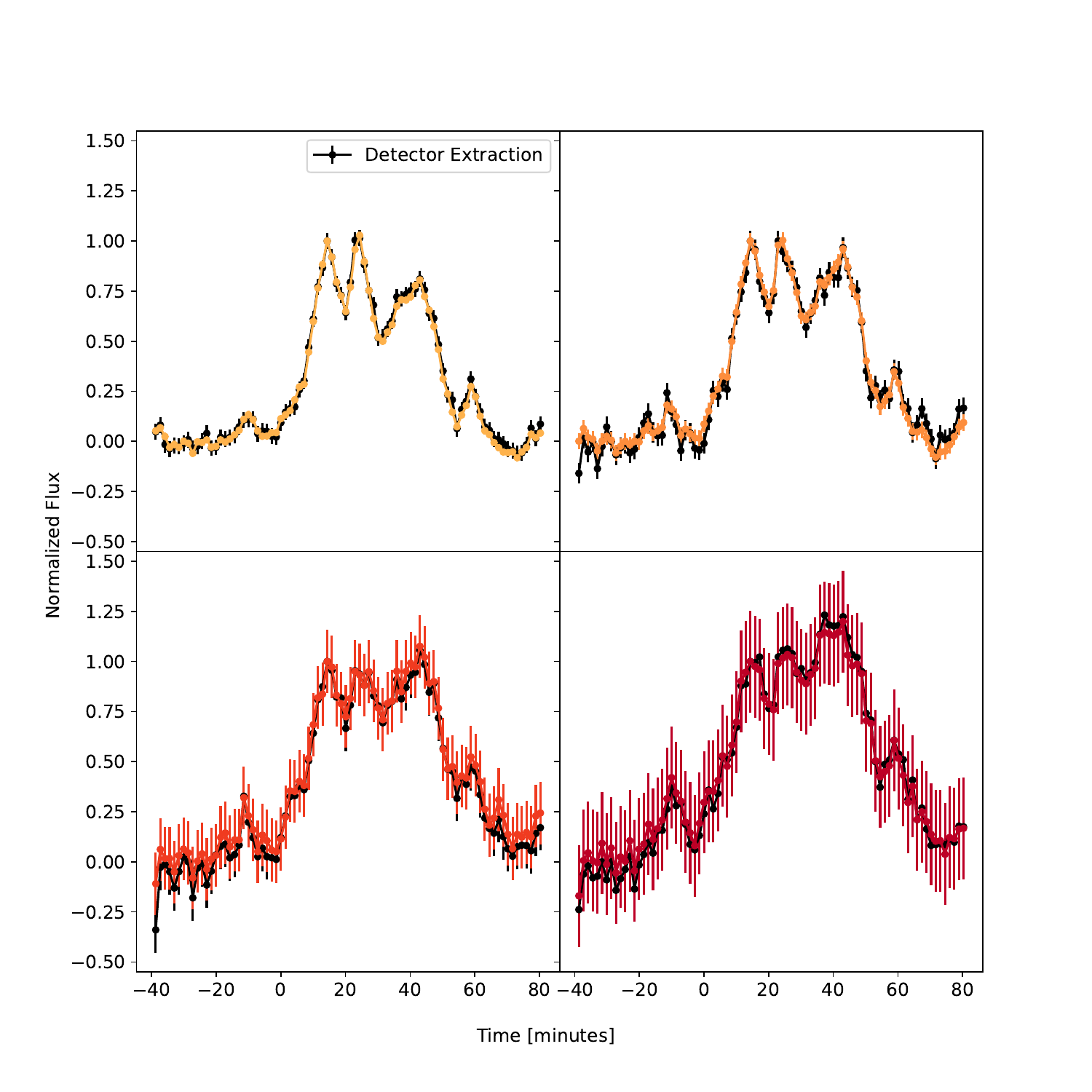}
    \caption{\textbf{Background-subtracted, detrended light curves extracted from 2D detector frames compared to cub-extracted light curves.} The colored light curves are the light curves used in the analysis of this paper. The black lines are the corresponding lines extracted from the 2D IFU detector, extracted from the detector slice corresponding to the \Sg\ spaxel in the 3D cube. Both light curves have been normalized using Eq.~\ref{eq: Flux Equation}.
    }
\label{fig:DetectorFrameLightcurve}
\end{figure}

\begin{table*}[ht]
    \caption{\textbf{Subtracted median flux densities, RMS values, and drift models}}
    \label{tab:reduction}
    \centering
    \begin{tabular}{lcccc}
        \hline \hline
        Channel& Median flux density [mJy] & RMS [mJy] & Drift [$10^{-3}$ mJy/h] & Drift error [$10^{-3}$ mJy/h]\\
        \hline
        Channel 1& \023.54 & 0.035& \013.4 & \01.6\\
        Channel 2& \092.67& 0.097 & \098.1 & \07.4\\
        Channel 3& 379.55& 0.309& \0\06.9 & 58.9\\
        Channel 4& 1100.1& 0.540 & 811.5 & 77.3\\
        \hline
    \end{tabular}
\end{table*}

To extract \Sg's flux density, we created median-subtracted data cubes. In the data cubes, \Sg\ is detected against the temporally constant background. We determined the pixel position of \Sg\ by fitting a circular Gaussian. The flux was measured using this single pixel. The flux density was normalized as
\begin{equation}
    F_{\rm{norm}}(t) = \dfrac{F(t) - F_{\rm{med}}}{F(t_0) - F_{\rm{med}}},
\end{equation}
where $F_{\rm{med}}$ stands for the median flux of each light curve $34~\mathrm{minutes}$ before the flare. We chose $t_0=12.49~\mathrm{minutes}$.
We estimated the uncertainty in the \Sg\ flux-density measurements by computing the standard deviation of the temporal variability in the reference-pixel light curves.  These are then propagated to the normalized light curve via Gaussian error propagation. In order to assess the impact of uncertainty of the flux by which we normalize the data, we compute 100 simulated normalization fluxes based on its error bar. The change in the spectral index is derived by fitting a power-law two the for channels, once for the real data with propagated uncertainties, and once for the set of 100 simulated normalized light curves. The change in the spectral index was determined by averaging the spectral index before and after the change, which gives a $\Delta \alpha=0.36$, the RMS based of the fitted and simulated error bar is $0.06$, which we round to $\Delta \alpha=0.4\pm 0.1$.
Table \ref{tab:reduction} gives the results for each channel. 

The pipeline-constructed 3D cubes use a complex algorithm (see \citealt{Law2023}) to assign the intensities measured on the 2D MRS detector to the corresponding spectral pixels (spaxels) in the 3D cubes. We verified that this procedure does not introduce photometric biases by measuring the light curves directly on the 2D detector pixels that correspond to \Sg's position at each wavelength. The resulting detrended light curves are shown in Figure~\ref{fig:DetectorFrameLightcurve}. These light curves have the same features as the light curves extracted from the 3D cubes and in particular show the same change in \Sg's spectral index.

\section{Comparison to previous MIR measurements}
The VISIR instrument at the VLT offers high sensitivity imaging at mid-infrared wavelengths, and in particular a filter centered on the PAH1 feature at $8.6\mathrm{\mu m}$.
Three works obtained flux limits with this instrument, \cite{Schodel2011} who studied the mean emission of Sgr~A* and obtained a $3\sigma$ flux density limit of $f_{3\sigma; \rm{mean}}=13.3~\mathrm{mJy}$, or $58.0~\mathrm{mJy}$ if an extinction value of $A_{8.6\mu m} = 1.6$ \citep{Fritz2011} is applied.
\cite{Haubois2012} obtained MIR observation during a bright NIR flare and obtained a $3\sigma$ flux limit of $5.1~\mathrm{mJy}$, or $22.4~\mathrm{mJy}$ if extinction correction is applied. The last work, \cite{Dinh2024}, focuses on compact objects and temperature maps of the central region and does not report a MIR flux density limit for \Sg.
Applying the same spectral selection as the PAH1 of VISIR, we obtain a PAH1 light curve, which we can directly compare against these observations. We obtain a peak flux density of $1.96~\mathrm{mJy}$ of the flare, corresponding to a de-extincted flux density of $8.6~\mathrm{mJy}$. We caution that the existing MIR extinction correction provided by \cite{Fritz2011} are based on much smaller aperture ISO/SWS observations \citep{Lutz1996}. The GC extinction is highly variable \citep[e.g.,][]{Fritz2011,Haggard2024}. Thus, the previously derived extinction laws may not be directly applicable for high spatial resolution JWST observations.
Most segments of the light curve are variable and seem to show intrinsic variation of \Sg. However, a detailed study of the MIR flux distribution is beyond the scope of this work. A conservative estimate on the lower limit of \Sg's flux density can be obtained with the RMS values of the measurement uncertainty (\autoref{tab:reduction}).

\section{Flare Modeling}\label{Apx:models}
\setcounter{figure}{0}
\setcounter{table}{0}
The accretion flow of \Sg\ is likely in a magnetically arrested state \citep{1974Ap&SS..28...45B,1976Ap&SS..42..401B,narayan2003}, based on EHT and GRAVITY observations \citep{eht_paper_V,GravityCollaboration2020_polarization_ale,EHT2024_polsgr} and simulations of the wind-fed accretion onto the Galactic center \citep{Ressler2020,Ressler2023}. In this scenario, a large amount of magnetic flux is accreted onto the black hole with the infalling gas. The flux on the horizon can then become strong enough to repel the accreting plasma in a flux eruption
\citep{2003ApJ...592.1042I,2008ApJ...677..317I,Tchekhovskoy_2011}. The ejection of flux occurs through the process of magnetic reconnection, which can potentially power high-energy flares from the region near the event horizon \citep{Ripperda2020,Ripperda2022}. The energies of particles and photons powered by the flare depend on the typical magnetization of the plasma near the event horizon \citep{2022PhRvL.129t5101C}, which is largely unconstrained for \Sg. The reconnection event produces a flux tube of vertical field (i.e., it transforms toroidal field into poloidal field) that can push away the accretion flow. Low-density plasma from the magnetospheric region feeding the reconnection layer will heat up due to the reconnection and populate the flux tube. Once the flux tube is ejected from the black hole magnetosphere, it orbits in the accretion disk \citep{porth2019,Dexter2020_flare}. It is protruded by Rayleigh--Taylor plumes due to the inward pointing gravity and the density contrast with the disk \citep{Ripperda2022,Zhdankin2023}. These Rayleigh--Taylor instabilities can accelerate the electrons in the flux tube to energies capable of powering infrared emission \citep{Zhdankin2023}. The Rayleigh--Taylor instabilities can also drive turbulence inside the cavity \citep{2007PhFl...19i4104S,Zhdankin2023} that can trap accelerated electrons \citep{Kempksi2023,Lemoine2023}.

The general idea of particle energization due to turbulence and reconnection in a magnetized and orbiting cavity motivated us to model the flare spectrum as arising from a spherical, orbiting, one-zone emission region with a constant magnetic field $B$ and constant radius $R_{\rm{flare}}$. A total of $N_\e$ electrons are injected into the region with a Gaussian temporal profile
\begin{equation}
    Q_{\rm{inj}} (t) = \dfrac{N_\e}{\sigma \sqrt{2 \pi }} \exp[-(t - t_{\rm max})^2/\sigma^2]\quad.
\end{equation}
The electrons' energy distribution is a power law given by:
\begin{align}
    \dfrac{dN}{d\gamma} & = \dfrac{-p - 1}{\gamma_{\rm{min}}} \left(\dfrac{\gamma}{\gamma_{\rm{min}}}\right)^{-p} :\quad \gamma_{\rm min}\le\gamma\le\gamma_{\rm max}\\ 
     & = 0 ~ :\quad\gamma < \gamma_{\rm{min}} \vee \gamma > \gamma_{\rm{max}}\quad.
\end{align}
Here $\gamma_{\rm{min}}$ and $\gamma_{\rm{max}}$ are the minimum and maximum electron Lorentz factors.
In the model, electrons cool continuously through their synchrotron emission with a cooling timescale given by Equation~\ref{eq:sync_cool}. The electron energy distribution as a function of time is given by the continuity equation 
\begin{equation}
    \dfrac{\partial N_\e(\gamma, t)}{\partial t} = Q_{\rm{inj}} - \dfrac{\partial (\dot{\gamma}N_\e)}{\partial \gamma}\quad.
\label{eq:cont}
\end{equation}
No particle escape term \citep{Blumenthal1970,Dodds-Eden2011} is included in the model as motivated by the long confinement timescale  \citep{Kempksi2023,Lemoine2023} 
\begin{equation}
    \tau_{\rm{conf}} \propto \dfrac{r_g}{c} \left(\dfrac{r_g}{r_{\rm{Larmor}}}\right)^{1/3}
\end{equation}
and absence of substantial outflows from the flux tube region as motivated by GRMHD simulations \citep{Ripperda2022}. We solved Equation~\ref{eq:cont} numerically and computed the resulting synchrotron emission from the emission region for each time step using the code \textsc{flaremodel} \citep{Dallilar2022}. 

Once the intrinsic emission is known, the model accounts for Doppler boosting by computing the Doppler factor
\begin{equation}
    D_{\rm{boost}}(R_\mathrm{flare}, \phi, \Omega) = \dfrac{1}{\gamma(1- \beta \cos \phi \cos
{\Omega})} 
 \times D_{\rm{grav}} (R_\mathrm{flare})\quad,
\end{equation}
where $\beta = 2 \pi  R_\mathrm{flare} {(4 \pi^2 R_\mathrm{flare}^3)}^{-0.5}$, $\phi$ is the observer inclination angle, $\Omega$ is the angular position (``longitude'') of the emission region in its circular orbit, and $R_\mathrm{flare}$ is in units of gravitational radii. In addition, we included the gravitational redshift $D_{\rm{grav}}(R_\mathrm{flare}) = {(1-1/R_\mathrm{flare})}^{-0.5}$ in the Doppler factor $D = D_{\rm{boost}} \times D_{\rm{grav}}$ and computed the observed flux as
\begin{equation}
    F'_{\rm{model}}(\nu) = D^3 \times F_{\rm{model}}(\nu / D)\quad.
\end{equation}
The resulting boosted light curve shows a sinusoidal modulation. Other than the gravitational redshift, we have ignored general relativistic effects. 
The  model MIR light curves were normalized identically as the JWST measurements. 

The model was fit using the MCMC code \textsc{emcee} \citep{emcee2013} using a $\chi^2$ likelihood that included $\chi^2$ from all four MIRI channels and from the SMA light curve. 
We used a standard setup of 32 {``walkers''} and started the sampling chain at a region visually perceived as a good fit. The chain ran for 1000 steps, but we discarded the first $500$ steps and used only every fifteenth sample to minimize correlation in the sampling.
The fit converged well, and Figure~\ref{fig:mcmc_posterior} shows the posterior corner plots.  Table~\ref{tab:bestfit} shows the derived median and the 16\% and 84\% quantiles for the model's free parameters.

\subsection{Magnetic field strength in the accretion disk---comparison to Event Horizon Telescope modeling}
Modeling of the Event Horizon Telescope (EHT) observations allows to infer the magnetic field strength in the accretion disk of \Sg. Such a calculation often invokes a one-zone model that assumes the emission comes from a uniform sphere with radius $R_{\mathrm{flare}}=5GM/c^2$ \citep{eht_paper_V}. The sphere is filled with a collisionless plasma and a uniform magnetic field oriented to $\theta=\pi/3$ with respect to the line-of-sight. The observed specific flux from this sphere is given as
\begin{equation} \label{eq: Flux Equation}
    F_{\nu} = \frac{4}{3}\frac{\pi r^3}{D^2}j_{\nu},
\end{equation}
where $D$ is the distance to \Sg\ (8.127 kpc), and the specific volumetric synchrotron emission coefficient is taken from  Equation~(72) of \cite{Leung2011}:
\begin{equation*} \label{eq: j_nu}
    j_{\nu} = n_\mathrm{e} \frac{\sqrt{2}\pi e^2 \nu_\mathrm{s}}{3 c K_2(1/\Theta_\mathrm{e}\mathrm{)}} \left[ \left(\frac{\nu}{\nu_\mathrm{s}}\right)^{1/2}+2^{11/12}\left(\frac{\nu}{\nu_\mathrm{s}}\right)^{1/6} \right]^2 \exp{\left(-\left(\frac{\nu}{\nu_\mathrm{s}}\right)^{1/3}\right)},
\end{equation*}
which assumes a thermal distribution of electrons. Here, $K_2$ is a modified Bessel function of the second kind,  $\nu_\mathrm{s} = 2\nu_\mathrm{c} \Theta_\mathrm{e}^2 \sin(\theta)/9$, where $\nu_\mathrm{c}=eB/(2\pi m_\mathrm{e}c)$ is the cyclotron frequency, and $\Theta_\mathrm{e}=k_\mathrm{B} T_\mathrm{e}/(m_\mathrm{e} c^2)$ is the dimensionless electron temperature. 

The plasma $\beta$ parameter is used to compare the magnetic and gas pressure:
\begin{equation}
    \beta = \frac{8\pi k_\mathrm{B}(n_\mathrm{e}T_\mathrm{e}+n_\mathrm{i}T_\mathrm{i})}{B^2}
\end{equation}
where $T_\mathrm{e}$ and $T_\mathrm{i}$ are the electron and ion temperatures, respectively. If the plasma is fully ionized,  the number density of ions and electrons should be equal ($n_\mathrm{e}=n_\mathrm{i}$). Additionally, due to the collisionless nature of the plasma, the ions and electrons can have differential heating, leading to a non-unity value for the ion--electron temperature ratio $R=T_\mathrm{i}/T_\mathrm{e}$. These assumptions lead to an equation for the electron number density:
\begin{equation}
    n_\mathrm{e} = \beta\frac{B^2}{8\pi} \frac{R}{k_\mathrm{B}(R+1)T_\mathrm{i}}\quad.
    \label{eq: ne}
\end{equation}

\begin{figure}[h]
    \centering
    \includegraphics[width=0.5\textwidth]{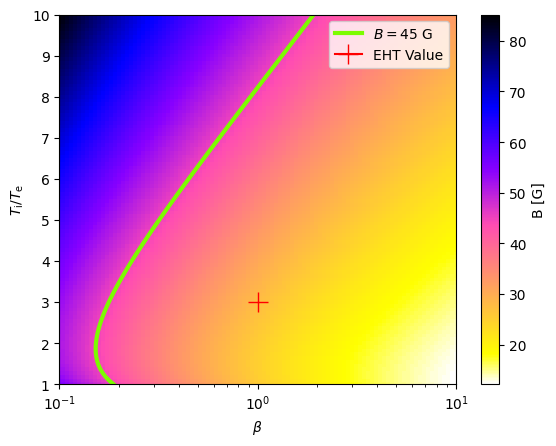}
    \caption{Distribution of model magnetic field values as $\beta$ and $T_\mathrm{i}/T_\mathrm{e}$ are varied. The green line shows all parameter combinations that result in a $45$~G magnetic field, while the red cross indicates EHT's magnetic field value for their chosen parameters of $\beta=1$ and $T_\mathrm{i}/T_\mathrm{e}=3$.}
    \label{fig: EHT Parameter Search}
\end{figure}

\begin{figure}
    \centering
    \includegraphics[width=0.65\textwidth]{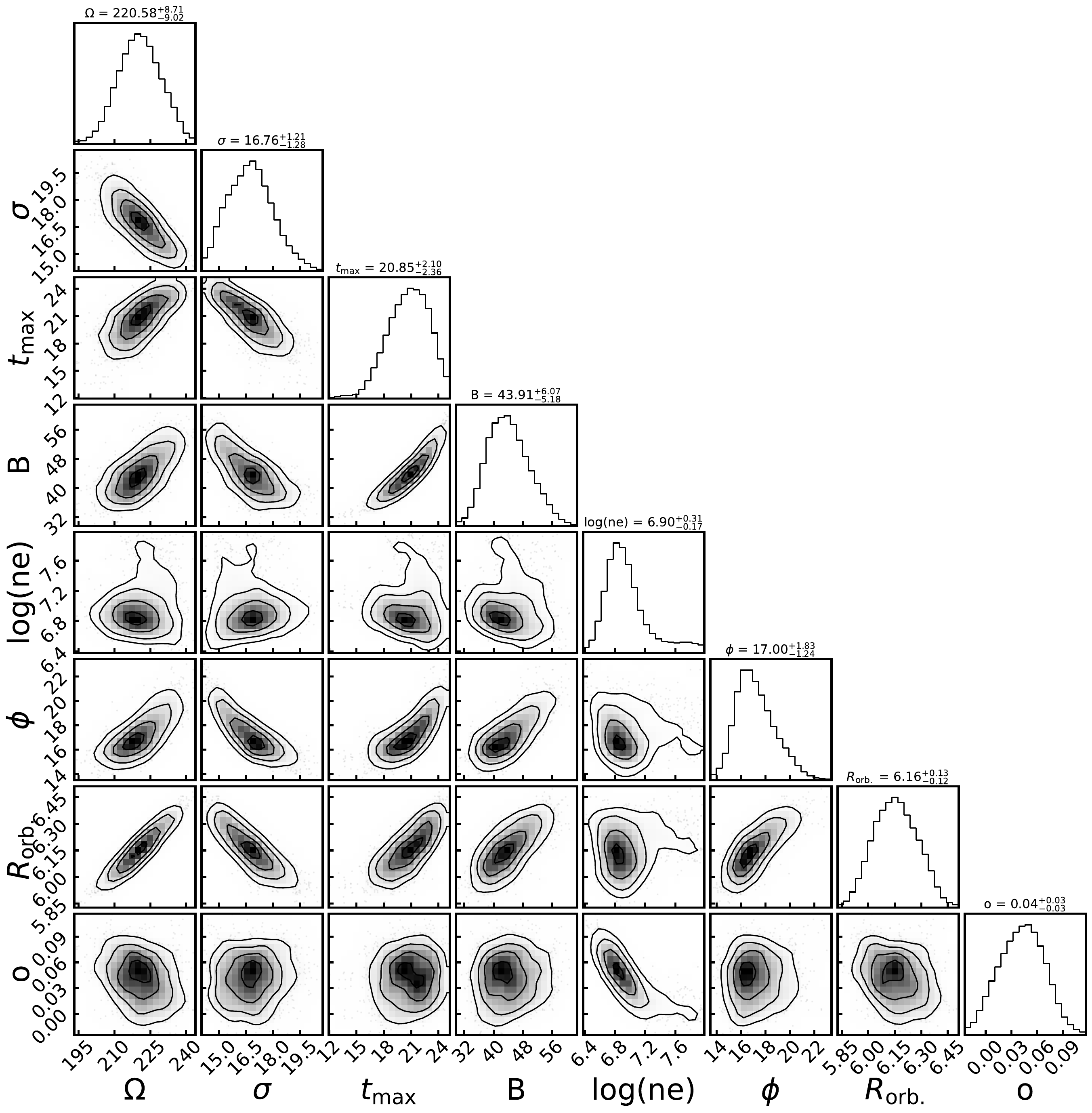}
    \caption{\textbf{Posterior of the MCMC chain.} Columns left to right show the starting angular position of the emitting region ($\Omega_0$), the width of the Gaussian electron-injection pulse ($\sigma$), the time of the maximum electron injection rate ($t_{\rm max}$), the magnetic-field strength ($B$), the total number of injected electrons ($\log(N_\e)$), the flux density of the constant 220~GHz component ($\phi$ in mJy), and the orbit radius of the emitting region in units of the Schwarzschild radius ($R_\mathrm{flare}/R_{\rm S}$). Histograms above or to the right of each column show the probability distribution for the corresponding parameter, and each parameter's median value and uncertainty (16\% and 84\% range) are shown above its histogram plot.}
    \label{fig:mcmc_posterior}
\end{figure}

The specific flux can be expressed in terms of only a few parameters by substituting $n_\mathrm{e}$ from Equation~\eqref{eq: ne} and $j_\nu$ from Equation~\eqref{eq: j_nu} into Equation~\eqref{eq: Flux Equation}. Assuming the ion temperature is a third of the virial temperature ($T_\mathrm{i} = GMm_\mathrm{p}/(9k_\mathrm{B}r) = 2.4\times 10^{11} \text{ K}$) and $F_{\nu}=2.4$ Jy from the 2017 ALMA campaign \citep{Wielgus_2022}, we are left with $R_\mathrm{flare}$, $\beta$, and $B$ as free parameters. We estimated the range of magnetic field strengths based on this one-zone model for a reasonable range of parameters by numerically solving for $B$ while varying $\beta\in[0.1,10]$ and $R_\mathrm{flare}\in[1,10]$. Figure~\ref{fig: EHT Parameter Search} shows that this results in a range of $B\in[12,85]$
which clearly includes both EHT's result of $B=30$~G \citep{eht_paper_V} as well as the fiducial magnetic field strength based on the MIR flare model, $B=45$~G. MAD simulations support $\beta$ values on the order $0.1$ in the the inner accretion region, where we expect the emission to originate \citep{Ressler2020,Ressler2023,Ripperda2022,Salas2024,Galishnikova2024}. Figure~\ref{fig: EHT Parameter Search} indicates that the larger magnetic field strengths estimated based on the MIR flare are within the range predicted based on the EHT observed submm emission.

\subsection{Model fits and parameter posteriors}
Certain parameters are only poorly constrained by the light curve data, and we had to fix them to reasonable fiducial values. 
One parameter is $\gamma_{\rm{min}}$, which we set to 10. This corresponds to the typically quoted energy of the ambient thermal electrons responsible for the bulk of the mm emission \citep[e.g.,][]{VonFellenberg2018}. The Chandra X-ray observations constrain $\gamma_{\rm{max}}$: depending on $\gamma_{\rm{max}}$ and the power-law slope $p$, the model could produce significant synchrotron flux at X-ray energies. The absence of the X-ray emission during the flare constrains $\gamma_{\rm{max}}$, and we set $\gamma_{\rm{max}}=3\times10^4$. 
Similarly, the normalization of the flux density leaves $p$ unconstrained. We therefore set $p=2$, which gives the canonically observed NIR spectral slope $F_{\nu,\rm{NIR}} \propto \nu^{-0.5}$ \citep[e.g.,][]{Hornstein2007}. 

\subsubsection{Single Injection Event}
Table \ref{tab:bestfit} reports the model's posterior parameters. The best-fit model has a reduced $\chi^2=3.5$. Because the light curve shows small but highly significant variations that the simple model cannot capture, we rescaled the error bars to obtain $\chi_r^2=1$. This required rescaling factors $f_{\rm{CH1}}=3.47$, $f_{\rm{CH2}}=1.35$, $f_{\rm{CH3}}=0.75$, $f_{\rm{CH4}}=0.7$, $f_{\rm{220GHz}}=1.3$. Unless otherwise stated, we adopted those rescaling factors for all fits described below.
The best-fit model has a magnetic field strength  $B=44_{-6}^{+5}~\mathrm{G}$ and $\log(n_\mathrm{e})=6.9^{+0.3}_{-0.2}$, resulting in $\sigma_{\rm{e}}\approx10$--50.

Changing $p$ to $p=3$ leaves $\Omega_0$, $\sigma$, $t_{\rm{max}}$, and $R_{\mathrm{flare}}$ unchanged, but the magnetic field strength increases to $52_{-5}^{+4}~\mathrm{G}$ and the electron density to $ 10^{7.1}$ cm$^{-3}$, resulting in very similar $\sigma_{\mathrm{e}} \sim 20$. The difference in $\chi^2$ is negligible ($\Delta \chi^2 = 0.1$, see Table~\ref{tab:fix_paramexplore}).

In the absence of astrometric or polarimetric measurements of the flare, the orbital parameters of the flare are poorly constrained. In particular, the inclination, which causes stronger or weaker magnification, is largely degenerate with the strength of the intrinsic emission. For the fit reported in Table~\ref{tab:bestfit}, both the inclination and the orbital radius are free parameters. The best-fit value of the inclination is $\phi=25^{\circ}\pm2^\circ$ and of orbital radius is $R_\mathrm{flare} \sim 6.6~\mathrm{R_\mathrm{S}}$. These values are consistent with observations by GRAVITY and ALMA and with the statistics of the NIR light curve \citep{GRAVITY_collab2023_polariflares, Wielgus2022, vonFellenberg2024}.

\begin{figure}
    \centering
    \includegraphics[width=0.5\textwidth]{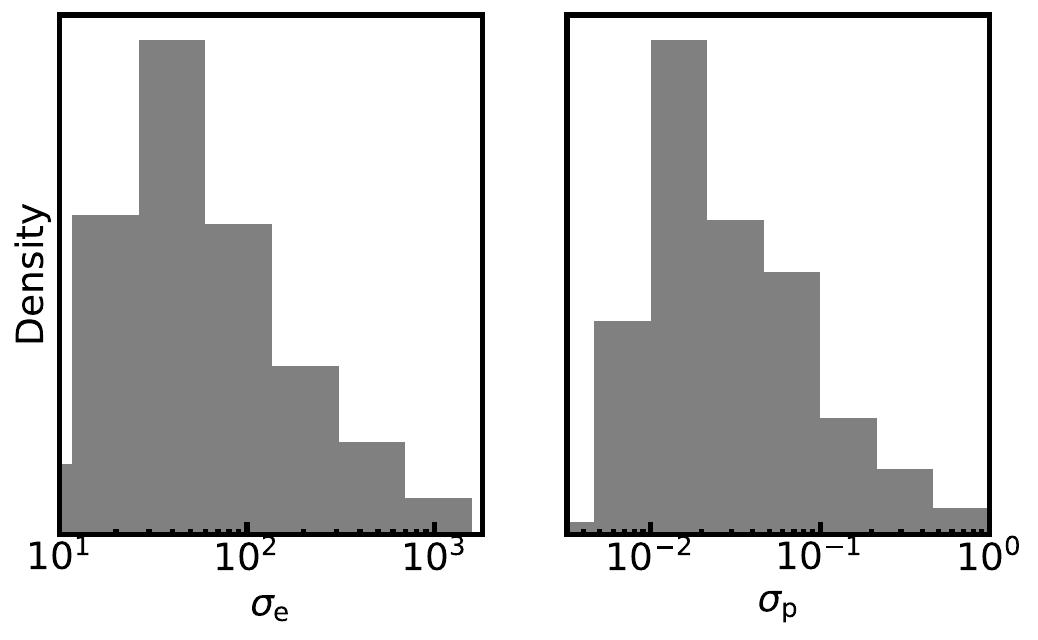}
    \caption{\textbf{Posteriors of the electron magnetization $\sigma_{\rm{e}}$ (left) and proton magnetization $\sigma_{\rm{p}}$ (right) of the flare.} 
    The thick vertical lines indicate the median values, and the dashed lines mark the quartiles.
    }
    \label{fig:magnetization}
\end{figure}

The electron magnetization $\sigma_{\rm{e}}$ depends on the magnetic field strength and the particle density. Because the luminosity of the flare depends on the size of the emission zone, $L \propto n_\mathrm{e}  R^3$, the magnetization is not directly constrained. In order to estimate the range of allowed electron magnetization, we adopted a flat prior distribution of allow emission regions $R_{\mathrm{flare}} \in  [0.7R_\mathrm{S},2.3R_\mathrm{S}]$,  motivated by the typical flux tube sizes in GRMHD simulations \cite{Ripperda2022}. For numerical reasons, we fixed the model parameters, which show only weak correlations with the magnetic field strength, and the electron density, $\phi$, $R_\mathrm{flare}$, and the submm offset $o$. In order to compute the posterior distribution of the magnetization, we used  \textsc{dynesty} \citep{Skilling2004_dynesty, 
Skilling2006_dynesty}. This fit derives a median posterior magnetic field strength of $46^{+7}_{6}~\mathrm{G}$ and $\log(n_\mathrm{e})=5.6_{-0.7}^{+1.1}$ (Table~\ref{tab:freeR}).  The posterior distribution is shown in Figure~\ref{fig:magnetization}. The median magnetization is $\sim$350 with lower and upper quartiles at $(110, 810)$. This range depends on the prior width chosen for the flare emission-region size $R_{\mathrm{flare}}$. 

\subsubsection{Double Injection Event}\label{sec:doubleGauss}
The observed flare shows a secondary peak at $t\approx20~\mathrm{minutes}$. The baseline model does not explain this peak, but several scenarios could explain it.  Examples include a variation in the magnetic field strength or a deviation from a purely Gaussian injection profile. We modeled the latter scenario by adding a secondary Gaussian injection event, slightly improving the fit ($\Delta \chi^2 = -0.5$).  Table~\ref{tab:doubleGauss} gives the fit posterior values. The magnetic field strength is slightly higher ($\sim$62~G) with correspondingly lower $\log(n_\mathrm{e})\sim6.0$, leading to $\sigma_{\rm{e}}\approx 420$ for $R_{\mathrm{flare}}=1~{R_\mathrm{S}}$.
Allowing for a secondary injection event is not the only explanation that could model the second peak in the light curve. Plausible alternatives include a variable magnetic field, a shorter orbital period for a spinning black hole, or a variable emission radius.

\subsubsection{Fit without sub-mm contribution}
We explored the possibility that the mm increase is by chance and unrelated to the observed MIR variability by fixing the electron number density to $n_\mathrm{e}=10^{5.5}~\mathrm{cm^{-3}}$. The resulting best fit, reported in Table~\ref{tab:no_submm}, is largely consistent with the fit reported in Table~\ref{tab:bestfit}. However, the magnetic field is slightly lower, $B\approx 38$~G. Based on the requirement to not produce submm flux, $\sigma_{\rm{e}}>280$. 

\subsection{Summary of fit results}
Depending on the choices of the model parameters, the electron magnetization is constrained to be within $B\approx40$--70~G and $\log(n_\mathrm{e}) \approx 10^6$--10$^7$, resulting in $\sigma_{\rm{e}}\approx 10$--900 if the submm variability is considered to be causally connected to the MIR variability. If this is not the case, i.e., the submm variability is generated by a separate process, the submm flux produced by the MIR flare has to be lower than the measurement uncertainty in the submm measurement. Assuming $\gamma_{\rm{min}}\equiv10$, this places a constraint on electron density and, given $\sigma_{\rm{e}}\propto n_\mathrm{e}^{-1}$,  $\sigma_{\rm{e}}\geq300$.
The statistical uncertainty on the model parameters is smaller than the differences between the models, indicating a large theoretical uncertainty due to the simplicity of the model. We therefore caution that a more rigorous treatment of the electron injection or a different treatment of the relativistic modulation of the light curve may alter the values.




\begin{table*}
    \caption{\textbf{Model Posteriors}}
    \label{tab:bestfit}
       \label{tab:freeR}
        \label{tab:fix_paramexplore}
         \label{tab:no_submm}
    \centering
    \renewcommand{\arraystretch}{2} 
    \setlength{\tabcolsep}{5.pt} 
    \begin{tabular}{p{30mm}@{\hspace{4pt}}@{\hspace{2pt}}c@{\hspace{1pt}}@{\hspace{2pt}}c@{\hspace{2pt}}@{\hspace{2pt}}|cccccccc}
    \hline\hline
         Model Posterior &    
         $p$& $R_{{\mathrm{flare}}}$ &  $\Omega_0$&$\sigma$ & $t_{\rm max}$ & 
         $\log(n_\e)$ &$B$  & $\phi$&$R_{\mathrm{orb}}$& $\chi_{\rm{red.}}^2$\\
         Type&&$[R_{\rm{S}}]$&$[^\circ]$&\multicolumn{2}{c}{[minutes]}&[cm$^{-3}$]&[G]&[$^\circ$]&$[R_{\rm{S}}]$\\
         \hline
         Best Fit & $2.0$&$1.0$&$205_{-5}^{+5}$&$16.5^{+0.6}_{-0.6}$& $19.6^{+0.7}_{-0.7}$&$6.7^{+0.1}_{-0.1}$&$39_{-3}^{+2}$&$20^{+1}_{-1}$ &$6.0^{+0.1}_{-0.1}$ &$3.5$\\
         & $2.0$&$1.0$& $221_{-9}^{+9}$&$16.7^{+1.3}_{-1.3}$& $20.9^{+2.1}_{-2.4}$&$6.9^{+0.3}_{-0.2}$&$44_{-6}^{+5}$&$17^{+2}_{-2}$ &$6.2^{+0.1}_{-0.1}$ &$\equiv1.0$\\
\hline
\multicolumn{2}{l}{Two~Electron-Injection~Phases:}&&\null\\
First & 2.0 & 1.0 & $217_{-8}^{+9}$&$16.1_{-1.1}^{+1.8}$&$25.1_{-0.6}^{+0.5}$&$6.0_{-0.1}^{+0.1}$\\
Second & 2.0 & 1.0 &$1.7_{-0.7}^{+0.4}$&$24.7_{-0.2}^{+0.2}$&$5.3_{-0.1}^{+0.1}$&\\
\hline
          Different Electron \newline Power-law\newline
          Distribution Slope & \textbf{3.0}& $1.0$ &$254_{-4}^{+7}$&$12.3_{-0.2}^{+0.2}$&$25.1_{-0.6}^{+0.7}$&$7.1_{-0.6}^{+0.3}$&$52_{-5}^{+5}$&30&$6.6_{-0.1}^{+0.1}$&$1.2$\\
\hline
Free Emission \newline Zone Radius &          2.0& ${\in} [0.8,2.3]$ &$252_{-4}^{+5}$ & $14.4_{-0.6}^{+0.6}$& $21.8_{-2.2.}^{+1.6}$& $5.8^{+1.1}_{-0.7}$& $47_{-7}^{+7}$&
20& 6.2 &\nodata\\ 
\hline
    No mm Flare &          $2.0$&$1.0$ & $188_{-11}^{+9}$&$21.4^{+1.9}_{-1.9}$& $15.9^{+2.9}_{-3.6}$&5&$38_{-4}^{+4}$&$14^{+1}_{-1}$ &$5.8^{+0.1}_{-0.2}$ &$0.9$\\
    \hline
    \end{tabular}
\end{table*}

\begin{table}
    \centering
    \caption{\textbf{Model Posteriors with Two Electron-Injection Phases\tablenotemark{a}}}
    \label{tab:doubleGauss}
    \renewcommand{\arraystretch}{2} 
    \setlength{\tabcolsep}{3pt} 
    \begin{tabular}{ccccccccccc}
    \hline\hline
         $\Omega_0 $&$\sigma_1$ & $t_{\rm{max}_1}$ & 
         $\log(n_{\e_1})$&$\sigma_2~ $& $t_{\rm{max}_2}$& 
         $\log(n_{\e_2})$ &$B$  & $o$&$R$& $\chi_{\rm{red}}^2$\\
         
                  $[^{\circ}]$&$[\mathrm{min}]$ & $[\mathrm{min}]$ & 
         \null&$[\mathrm{min}]$ & $[\mathrm{min}]$ & 
         \null &$[\mathrm{G}]$  & $[\mathrm{Jy}]$&$[\mathrm{R_g}]$& \null\\
         \hline
         $217_{-8}^{+9}$&$16.1_{-1.1}^{+1.8}$&$25.1_{-0.6}^{+0.5}$&$6.0_{-0.1}^{+0.1}$&$1.7_{-0.7}^{+0.4}$&$24.7_{-0.2}^{+0.2}$&$5.3_{-0.1}^{+0.1}$&$62_{-3}^{+3}$&$0.2_{-0.1}^{+0.1}$&$6.2_{-0.1}^{+0.1}$&$2.6$\\
         \hline
    \end{tabular}
    \vskip 5pt
\tablenotetext{a}{for a $p=2$, $i=30^{\circ}$, and $R_{\mathrm{flare}}=1.0R_{\rm S}$ model.}
\end{table}

\end{document}